\documentclass{aa}  

\makeatletter
\renewcommand*\@fnsymbol[1]{%
  \ensuremath{%
    \ifcase#1
    \or\star
    \or\dagger
    \or{\star}{\star}{\star}%
    \or{\star}{\star}{\star}{\star}
    \or\ddagger
    \or\dagger\dagger
    \or\ddagger\ddagger
    \or\mathsection
    \or\mathparagraph
    \or\|
    \or**
    \or\dagger
    \else\@ctrerr
    \fi
  }%
}
\makeatother

\usepackage{graphicx}
\usepackage{txfonts}
\usepackage{xcolor}
\usepackage[colorlinks=true,
    linkcolor=blue,
    filecolor=magenta,      
    urlcolor=blue,
    citecolor=blue]{hyperref}
\usepackage{lastpage}

\usepackage{lipsum}
\usepackage{subcaption}         
\usepackage{lscape}             
\usepackage{placeins}           

\newcommand{\Teff}{\ifmmode {T_{\rm eff}}\else${T_{\rm eff}}$\fi}
\newcommand{\Msun}{\ensuremath{M_\odot}\xspace}

\newcommand{\Zsun}{\ensuremath{Z_\odot}\xspace}

\newcommand{\posydon}{\texttt{POSYDON}\xspace}

\newcommand{\mesa}{\texttt{MESA}\xspace}

\newcommand{\orcid}[2][0000-0000-0000-0000]{\href{https://orcid.org/#1}{#2}}

\begin{document}

   \title{Metallicity dependence of Wolf-Rayet binaries using detailed binary models}

   \subtitle{An absence of long-period systems at low metallicity}

%
    \titlerunning{Metallicity dependence of Wolf-Rayet binaries using detailed binary models}
    \authorrunning{M.M. Briel et al.}

   \author{\orcid[0000-0002-6842-3021]{Max M. Briel} \inst{1,2}\fnmsep\thanks{E-mail: max.briel@gmail.com}
           \and
           \orcid[0009-0009-1888-8785]{Eirini Kasdagli} \inst{3}
           \and
           \orcid[0000-0003-1474-1523]{Tassos Fragos} \inst{1,2}
           \and
           \orcid[0000-0003-0642-8107]{Tomer Shenar} \inst{4}
           \and
           \orcid[0000-0001-5261-3923]{Jeff J. Andrews} \inst{3,5}
           \and
           \orcid[0009-0008-4869-2284]{Denzel Goh} \inst{1}
           \and
           \orcid[0000-0002-6064-388X]{Abhishek Chattaraj} \inst{3}\fnmsep\thanks{Co-author as a member of the \posydon core developer team.}
           \and
           \orcid[0000-0001-6692-6410]{Seth Gossage} \inst{6,7}\fnmsep$^{\dagger}$
           \and
           \orcid[0000-0003-1749-6295]{Philipp Srivastava} \inst{6,7,8}\fnmsep$^{\dagger}$
        }

    \institute{Département d’Astronomie, Université de Genève, Chemin Pegasi 51, CH-1290 Versoix, Switzerland 
    \and
    Gravitational Wave Science Center (GWSC), Université de Genève, CH-1211 Geneva, Switzerland 
    \and
    Department of Physics, University of Florida, 2001 Museum Rd, Gainesville, FL 32611, USA 
    \and
    School of Physics and Astronomy, Tel Aviv University, Israel 
    \and
    Institute for Fundamental Theory, 2001 Museum Rd, Gainesville, FL 32611, USA 
    \and
    Center for Interdisciplinary Exploration and Research in Astrophysics (CIERA), Northwestern University, 1800 Sherman Ave, Evanston, IL 60201, USA 
    \and
    NSF-Simons AI Institute for the Sky (SkAI),172 E. Chestnut St., Chicago, IL 60611, USA 
    \and
    Electrical and Computer Engineering, Northwestern University, 2145 Sheridan Road, Evanston, IL 60208, USA 
}

   \date{Received August XX, 2026}

 
  \abstract
   {Observations of Wolf-Rayet (WR) stars in binaries in the SMC and LMC suggest a preference for short period ($P\lesssim30$ days) orbits, with an apparent absence of WRs in long-period systems. The Galactic population does extend to longer periods but shows a deficit in long-period WR binaries compared to their progenitors, the O-star population. Across these populations a nearly constant binary fraction has been observed.}
   {We aim to characterize the population of WR binaries across a range of metallicity and determine the metallicity dependence of the formation of these systems, specifically focusing on the period distribution. We compare our model predictions against the observed samples of WR binaries in the SMC, LMC, and Milky way to test whether they can reproduce the observed period distribution and constant binary fraction.
   } 
   {We use detailed binary evolution models from \posydon to predict the population of WR binaries at 0.01, 0.1, 0.2, 0.45, and 1\,\Zsun, analyzing the resulting period distribution and formation channels, while comparing them against the observed population.}
   {We find that the absence of wider WR binaries in the SMC, and potentially in the LMC, can be explained by stable, Case-B mass transfer only partially stripping the donor star, combined with WR winds at low metallicity being insufficiently strong to strip the remaining envelope. As a result, the long-period peak from Case-B mass transfer, present at $\Zsun$, disappears at low metallicity. We additionally find that stable mass transfer (SMT) produces a short-period peak through Case-A mass transfer ($P\sim5{-}10$ days) that closely matches observations across metallicity.
   Furthermore, SMT and non-interacting systems are the dominant formation channels of WR binaries at all metallicities, with their relative contribution showing no metallicity dependence. This result implies that the SMT channel has the same metallicity-dependence as isolated WR formation. At the same time, we find that common envelope evolution primarily produces short-period ($P<1$ day) WR binaries with black-hole companions.}
   {}

   \keywords{}

   \maketitle
\nolinenumbers 

\section{Introduction}

Being a powerful source of ionizing radiation, Wolf-Rayet (WR) stars are an important contributor in the feedback of massive stars on their environment. WR stars are an observational classification of massive stars, exhibiting strong, broad emission lines associated with ionized helium, nitrogen, and carbon, driven by an optically thick stellar wind and high surface temperatures \citep{Wolf+67, Crowther+07, Shenar+26}. WR stars are generally divided into classical WR (cWR) stars and "main-sequence" WR stars \citep[WNh;][]{deKoter+97}.
The latter are thought to be core-hydrogen burning stars, producing a WR spectrum, while cWRs, which make up 90\% of all observed WRs \citep{Shenar+19}, are expected to be a late evolutionary stage of massive stars, partially or fully stripped of their hydrogen envelopes \citep{Schmutz+89, Hamann+06, Hainich+15, Shenar+16, Schootemeijer+18}. Based on the observed surface abundances, cWRs are further divided into WN (WNE and WNL), WC, and WO subclasses, which are thought to correspond to an evolutionary sequence \citep{Smith+96}. 

The removal of the hydrogen envelope is crucial in the formation of cWRs \citep{Castor+75, Grafener+11}, which can be achieved through self-stripping or mass transfer in a binary.
In the self-stripping scenario, also known as the Conti-scenario, \citep{Conti+75, Abbott+87, Smith+14, Shenar+26} massive stars remove their hydrogen envelope through strong stellar wind mass loss, exposing their helium-burning core, giving rise to the WR spectra.
The zero-age main sequence (ZAMS) mass required to reach this stage is model-dependent, but is typically around $20{-}30\Msun$ at solar metallicity and increases to over $\gtrsim40\Msun$ for the Small Magellanic Cloud \citep[SMC;][]{Shenar+20}.
The metallicity dependence originates from the stellar wind strength decreasing with metallicity \citep[e.g.][]{Vink+01, Vink+05}, requiring an initially more massive stars to self-strip at low $Z$. However, the role of episodic eruptive mass loss, which is currently poorly understood, may also play a significant role in self-stripping at lower metallicities \citep{Pauli+26}.

Alternatively, the envelope can be removed through mass transfer in close binaries, where the outer envelope is stripped by mass transfer onto a companion \citep{Paczynski+67, Vanbeveren+80, Wellstein+99}. The binary channel is further split into two sub-channels depending on the stability of mass transfer. In the first sub-channel, unstable mass transfer occurs, leading to a common envelope (CE) phase \citep{Ivanova+13} which results in significant orbital shrinkage. If a merger is avoided, a cWR binary with a short orbital period can be formed after the envelope of the donor star is ejected. Stable mass transfer (SMT), on the other hand, is expected to lead to a wider range of final periods \citep[e.g.][]{Petrovic+05a, vandenHeuvel+17}, depending on the initial masses, orbital configuration, and mass transfer efficiency.

The binary channel was traditionally thought to be largely metallicity independent, since stripping is driven by Roche lobe overflow rather than metallicity-dependent stellar winds. This theory naively predicts that at lower metallicities more WRs should be found in binaries.
However, \citet{Shenar+20} showed that one should not expect binaries to increase in influence with metallicity. Observations across the SMC (${\sim}0.2\Zsun$), Large Magellanic Cloud (LMC; ${\sim}0.45\Zsun$), and MW (\Zsun) support this, showing a constant observed binary fraction of ${\sim}40\%$ \citep{vanderHucht+01, Bartzakos+01, Foellmi+03,Foellmi+03a, Schnurr+08, Shenar+20, Dsilva+20, Dsilva+22, Dsilva+23, Schootemeijer+24}. 

\begin{figure}
    \centering
    \includegraphics[width=\linewidth]{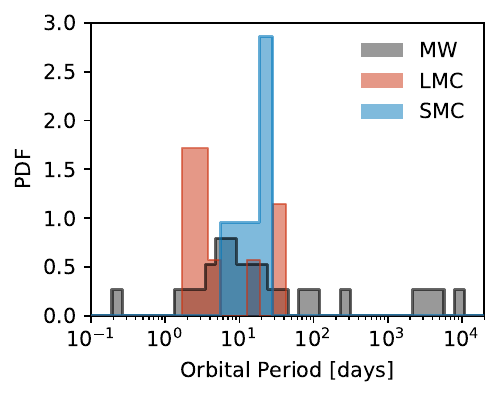}
    \caption{Normalised period distribution of observed WR binaries in the SMC (blue), LMC (red) and Milky Way (gray). The SMC observations come from \citet{Foellmi+03a, Shenar+16}, the LMC observations from \citet{Shenar+19}, and the Milky Way WR binaries from the post-mass transfer catalog from \citet{vanSon+26} which were originally compiled by \citet{vanderHucht+01}.}
    \label{fig:obs_period}
\end{figure}

At the same time, the WR stars found in binaries exhibit unexpected features in their orbital period distribution, as shown in Figure \ref{fig:obs_period}. The SMC WR population, thought to be complete, hosts 12 WR stars \citep{Neugent+18}, of which all five confirmed binaries exhibit periods smaller than 20 days \citep{Foellmi+03a, Shenar+16}. A modern high-resolution survey of this population by \citet{Schootemeijer+24} did not reveal new binaries in the sample, confirming a lack of long-period WR binaries in the SMC. The LMC shows a similar preference for short periods with known cWR binaries all having $P<40$ days \citep{Shenar+19}. While past surveys lose sensitivity at periods longer than a few months, a modern high-resolution survey by Shtainer et al. (in-prep) of the LMC WC/WO sample confirms the lack of long-period cWR binaries in the LMC.
In contrast, in the Milky Way, WR binaries cover a wider range of periods with systems up to several thousands of days, e.g. WR19 \citep[$P=10.1$ yrs][]{Veen+98, Williams+19} and WR140 \citep[$P=7.94$ yrs, e.g.][]{Moffat+86, Williams+19}. \citet{Dsilva+20, Dsilva+22, Dsilva+23} find a preference for $P<5$ days for Northern Galactic WN binaries and a peak at $P\sim5000$ days for Northern WC binaries. \citet{Deshmukh+24}, with a higher sensitivity to longer period binaries, does not find this difference between subtypes in the Southern Galactic cWR population, and attributes the initial discrepancy to possibly a limited sample size. They do find a stark contrast in the WR binary fraction at long-periods ($10^2-10^5$ days) where there is a lack of cWRs compared to massive O stars in binaries, indicating that interacting O stars in this regime might not experience a cWR phase or are removed from this period range. At the same time, they also find a lack of cWRs at periods beyond the regime of binary interaction, raising doubts on the effectiveness of the self-stripping scenario.

The observed period distribution of cWR binaries across metallicity raises questions about both the single-star and the binary formation mechanisms for WRs.
Previous population synthesis studies have attempted to reproduce the cWR population at SMC and LMC metallicity \citep{Pauli+22, Xu+25, Pauli+26}. For the LMC, \citet{Pauli+22} predict a preference for short periods around 5 days, which is longer than the observed periods of LMC cWR binaries, along with a significant contribution at $P>30$ days. For the SMC, \citet{Xu+25} similarly find a dominant short-period peak, but again accompanied by a non-negligible tail of long-period systems that are clearly absent in the observed populations. Both studies only consider cWR formation from the initially more massive stars due to their simulation setup, neglecting the contribution from the secondary to the cWR population. These results suggest that key formation pathways shaping the cWR period distribution at low metallicity might be incompletely modeled.

To address the open questions on the cWR period distribution, we present populations of cWR binaries across a wide range of metallicities using grids of detailed binary models, including the contribution of the secondary. We discuss the simulation setup in Section \ref{sec:methods}. We focus specifically on the resulting orbital period distribution and the formation channels responsible for producing WR stars in binaries, separating out formation through mass transfer and self-stripping. We show that SMT is a metallicity-dependent formation channel for WRs binaries, producing short-period WR binaries with a peak in period that closely matches the observed populations at low metallicities. Moreover, we find that low metallicity SMT fails to produce long-period WRs, offering a simple explanation for the absence of such systems in the SMC and LMC. We present these results in Section \ref{sec:results}, and discuss their robustness and implications in Section \ref{sec:discussion}, where we additionally discuss the differences with previous works. We summarize our conclusions in Section \ref{sec:conclusion}.

\section{Method} \label{sec:methods}

\subsection{Population synthesis}

We use the binary population synthesis code \posydon \citep{Fragos+23, Andrews+25}, which employs pre-computed grids of single- and binary-star models run with an adapted version of \mesa r11701 \citep{Paxton+11, Paxton+13, Paxton+15, Paxton+18, Paxton+19, Jermyn+23}, to produce synthetic stellar populations of cWR stars in binaries.
We use \posydon \texttt{v2.2.2}\footnote{Available on \href{https://github.com/POSYDON-code/POSYDON/releases/tag/v2.2.2}{Github}} with the grids from \texttt{Data Release 2}\footnote{Available on \href{https://zenodo.org/records/15194708}{Zenodo}} at $0.01\Zsun$, 0.1\Zsun, 0.2\Zsun, 0.45\Zsun, and \Zsun, with $\Zsun=0.0142$ \citep{Asplund+09}\footnote{$10^{-4}$, $10^{-3}$ and $2\Zsun$ metallicities are also available in Data Release, but not used in this work.}.
The \texttt{Data Release 2} contains five precomputed grids per metallicity, separated by evolutionary phase. Two single-star grids, \texttt{single\_HMS} and \texttt{single\_HeMS}, start their evolution at the ZAMS and as a fully-stripped helium star, respectively.
The three binary grids, \texttt{HMS-HMS}, \texttt{CO-HMS}, and \texttt{CO-HeMS}, cover the evolution of binary systems. 
The \texttt{HMS-HMS} grids start their evolution as two ZAMS stars, where the orbital evolution and stellar structure equations of both stars are solved with \mesa. In the \texttt{CO-HMS} and \texttt{CO-HeMS}, the evolution of a hydrogen-rich or a stripped star in a binary with  a compact object (CO) companion is followed. These precomputed binary grids include a self-consistent treatment of internal rotation and angular momentum transport, both within stellar interiors and between the stars and their orbit due to mass transfer, tides, stellar winds, etc. 

Because a detailed description of the stellar and binary physics is available in \citet{Fragos+23} and \citet{Andrews+25}, here we only summarize the physics assumptions that are essential in the formation of WR binaries. \posydon uses the stellar wind mass loss from \citet{Vink+00} for hot hydrogen-rich stars with a final metallicity dependence of $(Z/Z_\odot)^{0.68}$ \citep{Vink+01}. A fixed mass loss rate of $10^{-4}$ \Msun yr$^{-1}$ \citep{Belczynski+10a} is applied to very massive stars (luminous blue variables; LBVs) crossing the Humphreys-Davidson limit \citep{Humphreys+79}, which is defined in \posydon as stars having simultaneously $L>6\times10^5 L_\odot$ and $(R/R_\odot) \times (L/L_\odot)^{1/2} > 10^5$. For cool, red supergiant stars and asymptotic giant branch stars, a combination of the prescriptions from \citet{deJager+88}, \citet{Reimers+75}, and \citet{Bloecker+95} is used without a metallicity dependence. When a hot star ($T_\mathrm{eff} > 11,000\, \mathrm{K}$) reaches a surface hydrogen abundance below 0.4, the prescription by \citet{Nugis+00} is used with an explicit metallicity dependence. For more details on implementation of the stellar winds and their metallicity dependence in \posydon, see section 2.2.3 in \citet{Andrews+25}. 

For mass transfer occurring while the donor star is on the main-sequence, \posydon uses the \texttt{contact} scheme \citep{Marchant+16}, which allows for both stars to fill their Roche lobe simultaneously.
For interactions after the main-sequence, \posydon switches to the \texttt{kolb} scheme, which accounts for substantial radial expansion of the donor past the Roche lobe \citep{Kolb+90}. 
Since the mass transfer and subsequent detachment of the donor star is self-consistently modeled in \posydon, the donor star can detach before the full hydrogen envelope has been removed \citep[e.g.][]{Gotberg+17, Klencki+20}.
\posydon adopts the specific angular momentum accretion of \citet{deMink+13} and employs rotation-enhanced stellar winds to prevent the star from reaching critical rotation \citep{Paxton+15}.
Because tides are stronger in short-period systems, the accretor is able to dissipate angular momentum efficiently into the orbit and avoid spinning up to near-critical rotation, allowing it to continue accreting. 
As a result, the final mass transfer efficiency is higher (more conservative) in tight orbits than in longer-period systems \citep[e.g.][]{deMink+13, Sen+22, Rocha+24, Zapartas+25, Briel+26}.

While SMT is evolved within the grids of detailed binary simulations, unstable mass transfer is calculated on-the-fly following the $\alpha{-}\lambda$ energy formalism \citep[e.g.][]{Webbink+84, Livio+88, Ivanova+13}, when one of the instability criteria is reached in the \mesa model. For the exact conditions, see sections 4.2.4 and 4.1 in \citet{Fragos+23} and \citet{Andrews+25}, respectively.
For the common envelope evolution, inspired by 1-D hydrodynamic simulations \citep{Fragos+19}, we adopt a two-step formalism. First, the common envelope takes place for which we adopt $\alpha=1$ and compute the $\lambda$ structure parameter from the detailed stellar profiles at the onset of instability, using a core-envelope boundary defined at the hydrogen mass fraction of $X_H=0.3$. In the second phase, we then remove the remaining hydrogen envelope to $X_H=0.01$ through non-conservative stable mass transfer to expose the helium core \citep[for a more detailed explanation, see section 8.2 in][]{Fragos+23}.

We use the \citet{Fryer+12}-delayed supernova prescription with a mass-scaled natal kick drawn from a log-normal following \citet{Disberg+25}. We sample $3\times10^5$ binaries following a \citet{Kroupa+01} initial mass function between 7\Msun and 150\Msun and a flat mass ratio distribution. The samples follow a flat $\log_{10}(P)$ distribution \citep{Sana+25} between 0.75 day and 6000 days to cover the complete range of binary interactions.
We evolve the population to 50 Myr, after which no WRs are expected to form. For each star, we determine if there is a WR phase and track the properties of the system if so. Assuming a constant star formation history, we weight the period of each WR binary by the duration it spends at that period, including shifts in orbit as the system evolves.

\subsection{WR classification}

Since WR spectra arise from stellar winds that are sufficiently dense to become optically thick, we identify WR stars based on the optical depth ($\tau$), following the calculation in \citet{Langer+89}. We use a threshold of $\tau \geq 1.5$, which is calibrated against the observed threshold luminosity of WRs from \citet{Shenar+20} by \citet{Aguilera-Dena+22}. In Section~\ref{sec:discussion} and Appendix~\ref{app:optical_depth_boundary}, we show that our conclusions do not depend sensitively on the choice of this optical depth limit. We further sub-divide WR stars into three sub-classes: H-poor (WNL), H-free (WNE), and WC/WO based on their surface hydrogen ($X$) and ($Y$) mass fractions: 

\begin{itemize}
    \item \textbf{H-poor: } $0.05 \leq X_\mathrm{surface} \leq 0.5$ and $X_\mathrm{center} \leq 0.01$
    \item \textbf{H-free: } $X_\mathrm{surface} < 0.05$ and $Y_\mathrm{surface} \geq 0.7$
    \item \textbf{WC/WO: } $X_\mathrm{surface} < 0.05$ and  $Y_\mathrm{surface} < 0.7$
\end{itemize}
Although we distinguish between these three sub-classes, we do not analyze their individual sub-populations separately in this work and instead treat all three collectively as cWR stars as their period distributions are similar.

Observationally, cWR are separated from core-hydrogen-burning main-sequence WRs (WNh) based on their location in the Hertzsprung-Russell (HR) diagram and their surface hydrogen abundance. 
As in this work we are only interested in cWR binaries, we implement an additional cut based on the central hydrogen abundance, requiring $X_\mathrm{center}\leq 0.01$ for the \textbf{H-poor} class, which removes all core-hydrogen-burning systems from our sample. In Section \ref{sec:discussion}, we explicitly remove this criterion to show that it primarily affects non-interacting systems and does not significantly affect the main results of this work.

The upper limit for $X_H$ of 0.5 for the \textbf{H-poor} class is motivated by the high surface hydrogen fractions observed among WR stars in the SMC \citep{Hainich+15, Shenar+16}. For a WR to be considered H-free or WC, we adopt $X_\mathrm{surface}<0.05$, since this is of the same order of magnitude as typical observational abundance uncertainties \citep[see][]{Xu+25}. We note the exact choice of classification criteria affect the total number of WRs in the population, but does not significantly impact the resulting period distribution or its metallicity dependence.

\section{Results} \label{sec:results}

\subsection{Galactic WR population}

\begin{figure}
    \centering
    \includegraphics[width=\linewidth]{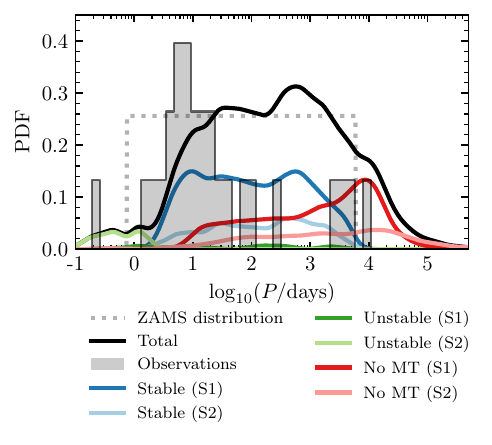}
    \caption{Period distribution of solar metallicty WR binaries per formation channel. The total (black line) integrates to 1 in $\log_{10}(P)$-space. The contribution of each channel is adjusted based on their relative contribution to the total population. The gray dotted line shows the initial period sampling of ZAMS binaries. The sample of Galactic WR binaries with known periods is shown as the grey density histogram which is rescaled with a factor 0.5 for comparison.}
    \label{fig:galactic_WR_period}
\end{figure}

Figure~\ref{fig:galactic_WR_period} shows the orbital period distribution of WR binaries in the $\Zsun$ population, separated by formation mechanism and whether the WR originates from the primary or secondary star in the binary. Despite the initial period distribution (gray, dotted line) being flat in $\log_{10}(P)$, the total WR period distribution (black, solid line) is shaped by binary interactions and has a general flat shape between ${\sim}20$ days and ${\sim}300$ days. The shortest period ZAMS binaries ($P_\mathrm{ZAMS}<1$ day) undergo unstable mass transfer and merge, thus are absent in the WR period distribution. The widest period systems above $10^4$ days are suppressed due to the initial sampling range up to $6000$ d, as discussed in more detail below; with sampling expanded to wider initial periods, a plateau from $10^4$ to the maximum sampled period is expected. We have included the observed Galactic WR binaries from \citet{vanSon+26} with known periods in Figure \ref{fig:galactic_WR_period} as the gray histogram \citep[see also the Galactic WR Catalogue][]{vanderHucht+01}. There is a mismatch between these observations, which prefer short periods, and our simulated galactic WR binary population, which provides a more uniform period distribution. We will discuss this in more detail in Section \ref{sec:galacic_obs}.

The darker, colored lines in Figure~\ref{fig:galactic_WR_period} indicate the primary star forming a WR. Initially primary stars are the more common sites of the WR phenomenon, as the primaries, by definition, are more massive. As such, they evolve first and are more likely to become stripped due to binary interactions, while their stronger winds make it easier to strip them.
As a consequence, the primary stars contribute ${\sim}70\%$ across all channels. Initially secondary stars contribute a non-negligible fraction (${\sim}30\%$) to the population, but this contribution ought to be sensitive to the physics governing the supernova, e.g., the natal kick and supernova remnant mass prescription.
Stable mass transfer is the dominant formation mechanism overall, with 41\% of primaries (dark blue) becoming a WR after interacting and with 14\% coming from secondaries (light blue) that underwent SMT before their WR phase. A substantial fraction (27\%) involves primaries that have not undergone mass transfer (dark red line), with a 11\% contribution from non-interacting secondaries (light red line).
Common envelope evolution contributes only a minor fraction to the total population (2\% for the primary and 4\% for secondaries). The secondaries populate a unique region in period, below $1$ day and are post-CE survivors with a BH companion which are likely to produce BBH mergers \citep{Belczynski+13}.

The SMT channel (dark blue) produces a weak bimodal distribution with peaks at $P\sim10$ days and $P\sim400$ days. These peaks originate, respectively, from the first mass transfer phase in the system being during the main-sequence (Case-A) and post-main sequence (Case-B), which were also found by \citet{Langer+20} for the O/B+BH population at LMC metallicity.
At the short orbital periods, the tides keep the companion synchronized to the orbit, preventing rapid spin-up during mass transfer \citep[see also][]{Sen+22, Briel+26}. As a result, the mass transfer is efficient with limited mass and angular momentum loss, leading to the orbit staying close to its ZAMS period throughout the interaction.
The longer-period peak instead originates from Case-B systems, where the mass transfer quickly spins up the companion to its critical velocity. This process is highly uncertain and depends strongly on the efficiency of transport of angular momentum inside the accretor and the specific angular momentum of the accreted material. As a result of the limited mass accretion, additional material is removed from the system. While this initially shrinks the orbit, the mass ratio is quickly switched and the orbit widens, leading to a small but noticeable decrease in systems between the Case-A and Case-B peaks (For more details see Section \ref{sec:Case_B} and Section \ref{sec:Case_A}). At sufficiently long periods, systems do not interact, leading to a decrease in WR systems beyond ${\sim}10^3$ from the SMT channel.

The second most common contribution in the \posydon galactic WR binary population is from non-interacting systems. The WR stars originating from the primary star (dark red line in Figure \ref{fig:galactic_WR_period}) consists of two sub-populations. 
The first is the continuum starting at $P\sim10$ days, which is produced by the highest-mass primaries, $\mathrm{M}_1\gtrsim90\,\Msun$ at $\Zsun$. These stars lose enough mass through stellar winds alone that they never expand to fill their Roche lobe, even at initial periods as short as $\sim2$ days \citep[see][]{Kruckow+24}.
For these initially massive stars the \citet{Vink+01} stellar wind prescription removes the envelope through self-stripping, as shown by figure 3 in \citet{Andrews+25} for the \posydon single stars. Thus their orbit widens substantially due to the strong wind mass loss during the main sequence, resulting in a minimum period during the WR phase of ${\sim}10$ days.

The peak/plateau at $10^4$ days instead originates from lower-mass primaries ($\mathrm{M}_\mathrm{ZAMS}\sim25{-}70\,\Msun$) with long initial periods ($P\gtrsim10^{3.5}$ days) that avoid interacting and undergo self-stripping. The 25\Msun lower limit represents the minimum mass at which models in \posydon self-strip at solar metallicity. The secondary contributes comparatively little to the no-mass-transfer channel, since the secondaries per definition have lower masses compared to the primaries and therefore less likely to have a mass within the range that leads to self-stripping. 
The primaries reach WR conditions through mass loss from main-sequence and LBV winds, but have less mass loss and orbital widening than the $\gtrsim90\,\Msun$ population.
These lower mass systems contribute more significantly to the period distribution due to the increased IMF weighting of lower mass stars.
While the exact contribution from each stellar wind mass loss prescription depends on the mass (and metallicity) of the primary, the higher the primary mass the more the main-sequence wind strips the star.
Since these long-period WR systems do not interact and are formed through self-stripping, the increased contribution from the $\mathrm{M}_\mathrm{ZAMS}\sim35{-}70\,\Msun$ population is expected to extend as a flat plateau towards wider periods where the two stars do not interact.
As such, the apparent drop above $\log_{10}(P/\mathrm{days})>4$ is an result of our initial sampling model.

WR stars formed through a common envelope are only a small fraction (3\%) of the total WR binary population.
In contrast to the other channels, the secondary star provides the dominant contribution, since the primary rarely survives the common envelope event with a stellar companion.
The secondary undergoes a common envelope with a BH companion to produce a distinct population with very short periods ($P<1$ day). These WR+BH systems populate a period regime where they are promising progenitors for gravitational wave mergers \citep{Belczynski+13} and X-ray binaries (see Section \ref{sec:galacic_obs}). However, the exact contribution of the CE channel is highly dependent on the assumed common envelope efficiency and structure parameter. However, the relative contribution of this channel to the WR binary population might provide constraints on common envelope physics, and warrant a detailed exploration beyond the scope of this work.

\subsubsection{Comparing with observations} \label{sec:galacic_obs}

The observational sample used in Figure \ref{fig:galactic_WR_period} is incomplete, but it nonetheless provides a useful benchmark for the range of periods that a realistic population must reproduce. While this incompleteness might introduce uncertainties, it shows distinct patterns across the orbital period regime. 

At the extreme short-period end, our model predicts a population of WR+BH systems with $P<1$ day, formed via the CE channel.
The Galactic WR binary Cygnus X-3, which has an orbital period of only 4.8 hours and is thought to host a compact object \citep{Ghosh+81, Singh+02, Bhargava+17}, likely originates from this channel.
Similarly, the extragalactic systems IC10 X-1 \citep{Prestwich+07, Wong+14} and NGC 300 X-1 \citep{Crowther+10, Binder+21}, with periods of 34.4 and 32.3 hours respectively, fall into this same short-period regime. Because Cygnus X-3 is currently the only known Galactic WR+BH candidate, the ratio of BH+WR systems to the total Galactic WR population can provide a unique method to probe the effects of the common envelope phase.

For the majority of the observed sample, the periods cluster around $P\sim6$ days, which coincides well with the Case-A SMT peak in the \posydon population. However, there is a notable dearth of observed systems between $P\sim200$ and $\sim2000$ days \citep[see also][]{Deshmukh+24}, where our model predicts a significant peak from the Case-B SMT channel. This discrepancy might be due to observational bias due to historical surveys primarily being sensitive to short-period systems \citep{Dsilva+20}, as the Galactic WR population has not received the same amount of careful analysis as the SMC and LMC WR populations have \citep{Shenar+26}. This mismatch might also point toward modeling uncertainty regarding stellar winds, which we discuss in Section \ref{sec:stellar_winds}.

Finally, at the longest observed periods ($P>2000$ days), the observations align with the plateau produced by the non-interacting $\mathrm{M}_\mathrm{ZAMS}\sim35{-}70\,\Msun$ progenitor population discussed above. The four observed systems with these long periods, WR~19, WR~125, WR~137, and WR~140, have eccentricities 0.80, 0.29, 0.31, and 0.89, and have O star companions, suggesting they have never gone through binary mass transfer.
According to \posydon, these wide systems are most likely formed through self-stripping rather than interaction, and the eccentricities of the observed systems are broadly consistent with this picture \citep{Massey+81}. 
On the other hand, mass transfer in eccentric systems does not guarantee circularization of the orbit \citep[e.g.][]{Rocha+24, Parkosidis+26}. For example, the presence of an OE star companion to WR~137 may indicate a more complex evolutionary history involving mass transfer \citep{Richardson+24}.
While the individual properties of these systems appear consistent with our model, their observed population density is lower than predicted in our model.
\citet{Deshmukh+24}, additionally, find a potential absence of WR binaries compared to massive O stars beyond the binary interaction limit.
Although their sample size is limited to 39 cWRs, this is in stark contrast with the predicted plateau of non-interacting systems at $P>10^4$ days in our model, suggesting that either the stellar winds in \posydon are too efficient at stripping single stars, or the effective binary fraction of $\sim30\Msun$ O/B stars is lower than expected from observations.

\subsection{The metallicity dependence of the WR period distribution} \label{sec:metallicity}

\begin{figure}
    \centering
    \includegraphics[width=\linewidth]{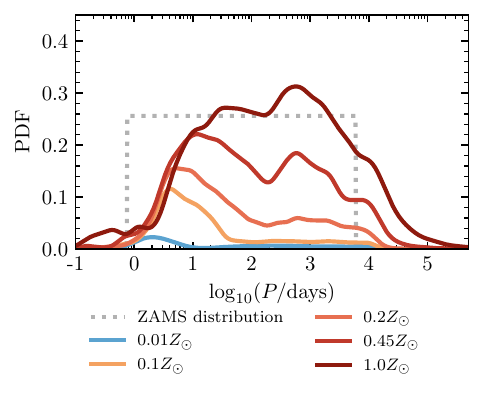}
    \caption{WR period distribution from the $0.01\Zsun$, $0.1\Zsun$, $0.2\Zsun$, $0.45\Zsun$ and $\Zsun$ populations. The PDFs at each metallicity are relative to the $\Zsun$ PDF, which means that the relative contribution of each metallicity is preserved, e.g. the $0.1\Zsun$ contributes less events than $\Zsun$. The ZAMS period distribution is shown as the gray dotted line.}
    \label{fig:metallicity_dependence}
\end{figure}

Figure \ref{fig:metallicity_dependence} shows the change of the period as a function of metallicity. Each distribution is normalized relative to the \Zsun probability density function (PDF), such that a lower overall WR binary rate at a given metallicity is directly reflected in a lower amplitude of the corresponding curve. We find that all sub-solar metallicities considered here are less efficient at producing WR binaries than \Zsun. This decreasing efficiency is accompanied by a systematic change in the shape of the period distribution. Towards lower metallicities, long-period WR binaries become increasingly disfavored. Opposite to what is naively expected from the binary channel, we find a clear decrease in the relative number of WR binaries with decreasing metallicity.

\begin{figure}
    \centering
    \includegraphics[width=\linewidth]{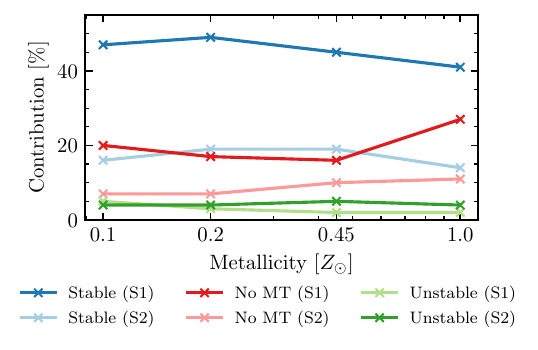}
    \caption{Relative contribution of each formation channel of WR binaries across metallicity. Although the \textit{No MT (S1)} contribution increases between $0.45\Zsun$ and $\Zsun$, at lower metallicities, the relative contribution of the SMT channels and non-interacting channels remains nearly constant across metallicity, indicating a similar metallicity-dependence in the formation of WRs through mass transfer as through self-stripping.}
    \label{fig:formation_channels}
\end{figure}

This decrease in the relative WR yield with decreasing metallicity cannot be attributed to a diminishing contribution from self-stripping alone. As shown in Figure \ref{fig:formation_channels}, the relative contributions of each sub-channel remain relatively constant as a function of metallicity. Since the non-interacting (self-stripping) formation channel becomes less efficient at lower metallicity, the constant relative contribution implies that the SMT channel must exhibit a similar metallicity dependence as the self-stripping scenario.

\begin{figure*}
    \centering
    \includegraphics[width=\linewidth]{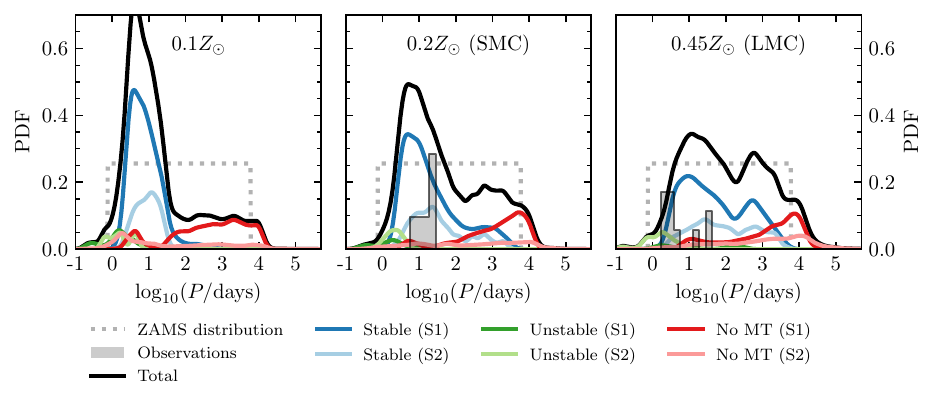}
    \caption{Period distributions of WR binaries at 0.1\Zsun, 0.2\Zsun (SMC-like metallicity), and 0.45\Zsun (LMC-like metallicity) from left to right. The colored lines indicate different formation channels of the WR binary. The gray dotted line indicates the ZAMS period distribution. For the SMC and LMC, we include observed WR binaries from \citet{Schootemeijer+24} and \citet{Shenar+19}, respectively. We have rescaled the observations by a factor of 0.1, because otherwise the probability density histogram of the observations falls outside the figure range.}
    \label{fig:SMC_LMC_period_dist}
\end{figure*}

To understand the reason for the change in shape of the WR period distribution as metallicity decreases, we show the WR period distribution PDFs at three difference metallicities (0.1\Zsun, 0.2\Zsun and 0.45\Zsun) and the different formation channels in Figure \ref{fig:SMC_LMC_period_dist}. The PDFs in this Figure are normalised to 1, but their relative contribution decreases with metallicity as shown in Figure \ref{fig:metallicity_dependence}.

The decrease in WR formation in binaries is primarily driven by the higher primary mass needed to produce WR winds at lower metallicity \citep{Shenar+20}, rather than by any changes in relative contribution of the formation channels, as Figure \ref{fig:formation_channels} shows. The non-interacting and SMT channels both produce fewer systems in absolute terms as metallicity decreases.
The decrease in non-interacting systems can be seen in Figure \ref{fig:metallicity_dependence}, where the bump at $10^4$ days shrinks towards lower metallicity and disappears at $0.1\Zsun$, as can also be seen in the left panel in Figure \ref{fig:SMC_LMC_period_dist}. At this metallicity and below, only the flat contribution from high-mass non-interacting populations remains, though their minimum period has increased from $10$ days at $\Zsun$ to $100$ days at $0.1\Zsun$. Due to reduced mass loss, these massive stars expand at lower $Z$ causing the binaries to interact at longer initial periods.

At the same time, the absolute contribution from SMT also decreases with metallicity with a reduced contribution of Case-A ($5$ days) and Case-B ($P\sim500$ days) SMT systems.  At $0.1\Zsun$ and $0.01\Zsun$, the Case-A SMT population is the only contribution to the SMT formation channel and the peak in period distribution has shifted down from ${\sim}10$ days at $\Zsun$ to $2$ days at $0.01\Zsun$ as can be seen in Figure \ref{fig:metallicity_dependence}. This is further confirmed by Figure \ref{fig:SMC_LMC_period_dist}, where as we move from the right panel to the left panel, the SMT peak from Case-A systems becomes more peaked at lower $P$ values. At $0.1\Zsun$ SMT only produces a sharp peak around $4$ days. Thus, as metallicity decreases, the Case-A SMT becomes the dominant interaction mechanism within the SMT channel to form WRs in binaries with Case-B systems no longer contributing at $Z\leq\,0.1\Zsun$.

The disappearance of the Case-B (and of longer-period Case-A) SMT systems is a result of a change in efficiency of envelope removal before core collapse. As metallicity decreases, the combination of SMT and weaker WR winds becomes insufficient to fully strip the hydrogen envelope. This effect is strongest for Case-B systems, which typically undergo a single mass transfer phase and leave behind a small part of the hydrogen envelope, independent of the metallicity. Because the orbits of these systems are sufficiently wide to prevent further interaction, the residual hydrogen envelope has to be removed through stellar winds. As such, a star that would become a WR star at $\Zsun$ becomes a partially stripped star at lower metallicities. We discuss this in more detail in Section \ref{sec:Case_B}. 
Case-A systems, on the other hand, experience two mass transfer phases, one during the main-sequence and a second post main sequence interaction. The first interaction already strips part of the outer layers, while the second interaction removes more of the envelope (for more details, see Section \ref{sec:Case_A}). The remaining hydrogen can more easily be stripped of by stellar winds compared to a Case-B system, allowing the donor star, even at 0.1\Zsun to reach a WR phase, where stellar winds alone would be too weak to do so. 

The metallicity dependence of the SMT and non-interacting channel are a consequence of the metallicity-dependence of the WR winds, which leads to an overall decreases in WR binaries with metallicity.
Furthermore, the relative contribution of each formation channel at each metallicity is nearly constant, as shown in Figure \ref{fig:formation_channels}.
The constant relative contributions and decreasing overall rate indicate that the SMT and non-interacting channel have similar metallicity dependencies, which implies that the contribution of WR binaries to the total population of WRs should remain similar across metallicity, since the non-interacting channel is effectively an isolated single star. This means that a constant binary fraction of WRs across metallicity is expected, which is as observed \citep{Shenar+20}, though the initial population properties as well as the contribution from mergers and runaways might affect the exact binary fraction.

\subsection{SMC and LMC populations} \label{sec:SMC_LMC}

The WRs in the SMC and LMC have extensively been searched for binarity \citep{Bartzakos+01,Foellmi+03, Foellmi+03a, Schnurr+08, Shenar+16, Shenar+19, Schootemeijer+24}, and only ${\sim}40\%$ of WRs have been found to have a companion.
These samples show a clear absence of long-period WR binaries with SMC and LMC WR binaries having periods of $P<20$ days and $P<40$ days, respectively \citep{Shenar+16, Shenar+19, Schootemeijer+24}. Figure \ref{fig:SMC_LMC_period_dist} shows these observations as the gray histogram in the middle and right panels for SMC-like and LMC-like metallicities. The SMC and LMC WR binaries have similar periods as the main peak originating from Case-A SMT in Figure \ref{fig:SMC_LMC_period_dist}.

While the SMC population of 12 WRs is thought to be observationally complete, the LMC WR sample of 154 stars is less well probed with modern surveys for binarity \citep{Neugent+18}. As such, the absence of long-period WRs above $P=20$ days for the SMC can also be used as a constraint on the predicted WR period distribution, while the absence of WR binaries above $P=30$ days for the LMC should be considered more carefully.
In the \posydon populations, contributions at long orbital periods are present, which originate from non-interacting and Case-B SMT systems. Especially at the LMC-like metallicity (right panel in Figure \ref{fig:SMC_LMC_period_dist}) the Case-B SMT contributes significantly.
However, as discussed in Section \ref{sec:metallicity}, the Case-B SMT contribution disappears at $0.1\Zsun$ due a combination of partial stripping and weaker stellar wind mass loss. The absence of long-period WRs indicates the requirement for weaker stellar winds and possible increased partial stripping at $0.2\Zsun$ and $0.45\Zsun$.
At the \posydon population at $0.1\Zsun$ indicates, these effects create a strong preference for short-period WR binaries formed through SMT and an much lower likelihood of forming long-period WRs binaries, which aligns more closely with the observed samples in the SMC and LMC.

\section{Discussion} \label{sec:discussion}

\subsection{Wolf-Rayet selection} \label{sec:other_WR_selection}

\begin{figure*}
    \centering
    \includegraphics[width=\linewidth]{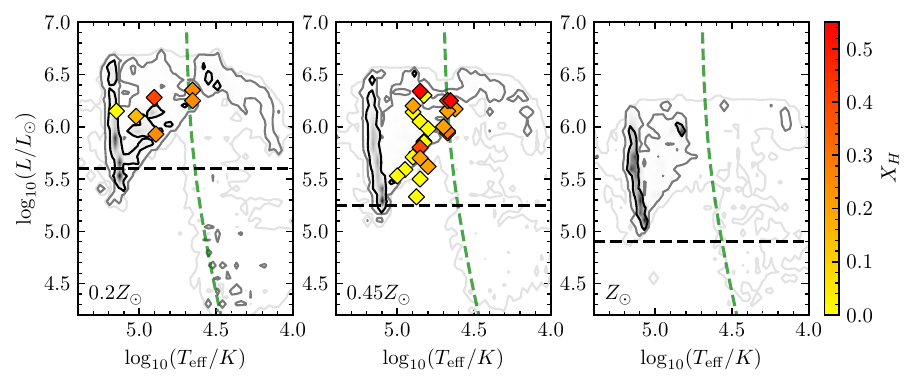}
    \caption{Hertzsprung-Russell diagrams at SMC-like (0.2\Zsun), LMC-like (0.45\Zsun), and MW-like ($\Zsun$) metallicities of the WR binary population. The gray colourmap indicates the WR binary population with the three contours indicating the 65\%, 95\%, and 99\% weight of the population. The green dashed line indicates the ZAMS, while the black dashed line indicates the observational lower luminosity limited for WRs in the SMC, LMC, and MW from \citet{Shenar+20}. Although the 99\% contour contains systems outside the expected region for WRs, the majority of systems can be found left of the ZAMS as expected from observations.}
    \label{fig:HR_diagrams}
\end{figure*}

Identifying the theoretical stellar model parameters that define the WR observational class remains a significant challenge. Without the use of stellar atmosphere modeling, it is difficult to accurately determine synthetic populations of WR stars \citep[see][for an example of self-consistent modeling]{Groh+14}. We, therefore, implement alternative selection criteria to determine the influence of the selection criteria on our results. 

\subsubsection{HR diagram location}
To asses the current selection criteria, we first compare the synthetic populations to observations. Figure \ref{fig:HR_diagrams} shows Hertzsprung-Russell diagrams of the \posydon synthetic WR binaries at $\Zsun$, $0.45\Zsun$, and $0.2\Zsun$. In these two-dimensional colormaps, the WR binaries are plotted with contour boundaries at 65\%, 95\% and 99\% of the population. At all three metallicities, the majority of WR binaries are located in the region where WR stars are observationally expected: to the left of the ZAMS, indicated by the dashed, green line. While a small part of the population appears to the right of the ZAMS within the 99\% contour, this contribution is negligible and should not significantly impact our results.

Furthermore, we compare our models to the observationally inferred minimum luminosity of WRs from \citet{Shenar+20}, which are $\log(L/L_\odot)=$ 4.9, 5.25, 5.6 for the \Zsun, LMC, and SMC populations, respectively. Although the optical depth limits from \citet{Aguilera-Dena+22} are calibrated to these boundaries using stripped single stars, the bulk of the \posydon WR binaries also lie above these lower limits at each metallicity. In the SMC and LMC, a small fraction of the population falls below this limit. This might suggest a requirement for weaker winds at lower metallicity, though the low number of WRs in the SMC makes its boundary sensitive to low-number statistics. Additionally, the use of \citet{Nugis+00} for helium star winds may overestimate mass-loss rates for lower-mass stars, potentially leading to their inclusion in the WR population (see Section \ref{sec:stellar_winds} for more details).

\subsubsection{Selection criteria}

\begin{figure*}
    \centering
    \includegraphics[width=\linewidth]{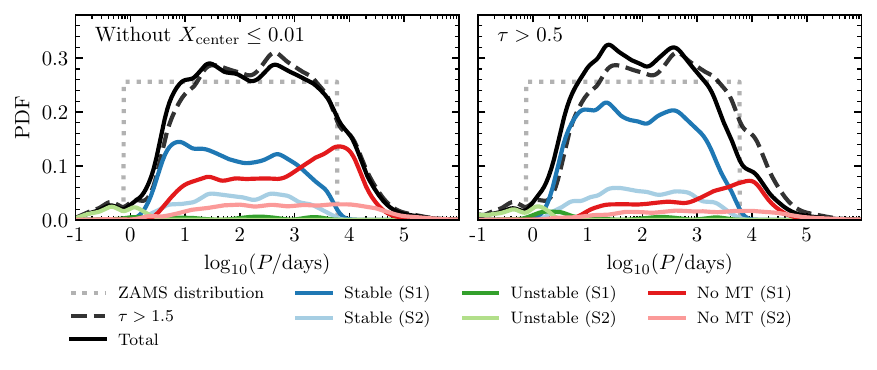}
    \caption{Period distribution for altered selection criteria at $\Zsun$. The left panel show the WR period distribution when no theoretical selection on core-hydrogen burning is done, while the right panel shows the WR period distribution if the optical depth limit is set to $\tau=0.5$. The dashed line shows the total period distribution with the $\tau=1.5$ selection criteria.}
    \label{fig:selection_criteria}
\end{figure*}

To verify if our results are not affected by the selection criteria, we first examine the ``H-poor'' sub-type classification by removing the requirement for core-hydrogen exhaustion ($X_\mathrm{center} \leq 0.01$).
Observationally, it is not possible to assess if a WR star is core-hydrogen burning, but selection criteria on observations are applied to try to remove these based on their position in the HR diagram and surface hydrogen abundance. By removing this constraint, we allow our cWR population to be polluted by main-sequence WR stars. The left panel in Figure \ref{fig:selection_criteria} shows the population without core-hydrogen burning for \Zsun. Compared to the normal WR period distribution in Figure \ref{fig:galactic_WR_period}, the high-mass contribution from non-interacting WRs has increased, which are massive stars that reach the WR conditions on the main sequence. 

Secondly, we consider the impact of the optical depth boundary. While we adopt an optical depth boundary of $\tau=1.5$, recent work by \citet{Sen+23}, \citet{Pauli+26} and \citet{Kasdagli+26} suggest a boundary of $\tau=0.5$. This decrease in the optical depth boundary is shown in the right panel in Figure \ref{fig:selection_criteria}. It leads to a significant boost in the contribution of SMT WR stars from the primary star to 58\% of all WR binaries. Although the choice in selection criteria can affect the relative contributions of the formation channel, it does not drastically alter the shape of the period distribution. Furthermore, in Appendix \ref{app:optical_depth_boundary}, we show the WR period distribution across metallicities with the optical depth boundary set at $\tau=0.5$. It shows the same metallicity dependence as the $\tau=1.5$ boundary, indicating that the results in this work do not depend strongly on the exact optical depth criteria.

\subsection{Stellar Winds} \label{sec:stellar_winds}

The stellar-wind mass-loss rates in the \posydon models do not reproduce the WR population exactly, but show the effect that main-sequence, red supergiant (RSG), and WR winds have on the WR binary population and their period distribution.
At solar metallicity, \posydon predicts a long-period ($P>10^4$ days) plateau in the WR period distribution from $\mathrm{M}_\mathrm{ZAMS}=35{-}70\Msun$ non-interacting binaries. This long-period plateau is not observed in the Galactic WR population \citep{Deshmukh+24}. This discrepancy could originate from the limited sample of WR binaries, rather than a detection bias, as the interferometric work by \citet{Deshmukh+24} is not biased against long periods within their sensitivity domain.

Alternatively, \posydon may form WRs too efficiently from stars with $M_\mathrm{ZAMS}\gtrsim25\Msun$. Since most of the plateau WR binaries have lost the majority of their mass through a combination of main-sequence and RSG winds, these stellar-wind mass-loss rates might be overestimated in \posydon.
Both the RSG and main-sequence wind prescriptions in \posydon show possible signs of being too strong. Although the RSG winds are thought to be metallicity independent \citep[e.g.][]{Antoniadis+25}, supernova progenitor studies favor weaker RSG winds than \posydon currently assumes \citep{Zapartas+25a}.
For main-sequence winds, the \citet{Vink+01} prescription used for hot massive stars is known to overestimate the mass-loss rate by a factor of 2 to 3. These reductions in stellar winds could decrease the formation efficiency of WR binaries from the $\mathrm{M}_\mathrm{ZAMS}=35{-}70\Msun$ non-interacting binaries, particularly at long periods. Since the long-period plateau disappears towards lower $Z$ due to less efficient WR formation, a reduction in main-sequence wind mass loss at $\Zsun$ is likely to supress the plateau in a similar way.

\posydon predicts a second shorter-period plateau ($10$-$10^4$ days) also from non-interacting systems from massive stars experiencing strong mass loss from main-sequence winds at solar metallicity. At lower metallicities, the mass loss of these massive stars is dominated by the LBV winds in \posydon which are implemented with a fixed, metallicity-independent mass-loss rate following \citet{Belczynski+10a}. This might cause an artificial metallicity independence of the non-interacting systems of the WR plateau at low $Z$, e.g. the dark red line in the left panel of Figure \ref{fig:SMC_LMC_period_dist}. However, \citet{Pauli+26} shows that a more physically-motivated stellar wind mass loss prescription for LBV stars (near the Eddington-limit) leads to only a small metallicity-dependence, though does not address the absence of such systems at long periods. As such the effect of a fixed LBV mass loss might be less than expected.
In our models, the WR binaries from these massive stars are present at all explored metallicities across a large range of periods. For example, at $0.1\Zsun$, they are the dominant formation mechanism of WR binaries with $P>100$ days. An observational presence or absence of WR binaries in that regime woudl allow them to be used as a calibrator for very massive star wind mass loss.

\posydon uses the \citet{Nugis+00} wind prescription for WR stars, which was calibrated to observations. While this prescription is likely appropriate for high-mass WRs, it is known to overestimate the wind mass loss of lower-mass helium stars \citep[e.g.][]{Drout+23}. The impact of adopting a more appropriate (lower) wind mass loss rate for these stars in \posydon is illustrated in the appendix of \citet{Chattaraj+26a}. 
The overestimation of stellar wind mass loss leads to a higher optical depth for these low-mass helium stars, possibly pushing them into the $\tau>1.5$ regime. This contamination is expected to affect all metallicities similarly and its effect on the population should be comparable to shifting the $\tau$ boundary, as discussed in Section \ref{sec:other_WR_selection}.

\subsection{Partial stripping Case-B} \label{sec:Case_B}

\begin{figure}
    \centering
    \includegraphics[width=\linewidth]{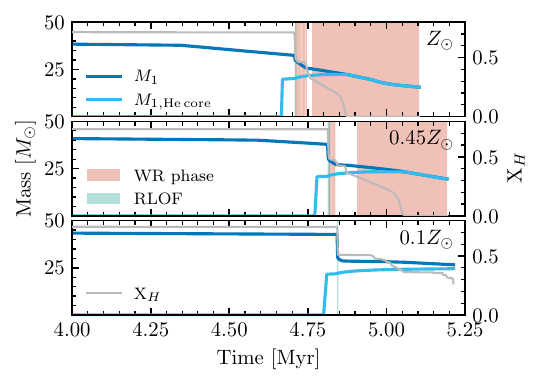}
    \caption{Evolution of an example model undergoing Case-B SMT with $P_\mathrm{ZAMS}=373$ days, $\mathrm{M}_1=45\,\Msun$, and $q=0.8$ at $\Zsun$ (top panel), $0.45\Zsun$ (middle panel) and $0.1\Zsun$ (top panel). The dark and light blue line shows the total primary star mass and the helium core mass, respectively. The duration of the Roche lobe overflow is indicated with the pink area. Due to the thermal timescale of the mass transfer, it appears near instantaneous in this figure. The orange area is when the model falls within the WR criteria, as discussed in Section \ref{sec:methods}. The criteria can be reached right after the mass transfer, as the star finds its thermal equilibrium, but does not contribute to the WR population. The $0.1\Zsun$ model does not experience a WR phase. The gray line and right y-axis shows the surface hydrogen abundance.}
    \label{fig:example_case_B}
\end{figure}

In Section \ref{sec:metallicity}, we attribute the absence of long-period WR binaries from SMT as an effect of partial stripping and reduced stellar winds strength. In this section, we show a representative example binary undergoing Case-B mass transfer to show that the combination of stellar winds and partial stripping causes the absence of WRs from Case-B SMT systems at low metallicity.

The binary in Figure \ref{fig:example_case_B} has the initial properties of $P_\mathrm{ZAMS}=373$ days, $\mathrm{M}_1=45\,\Msun$, and $q=0.8$, and is shown at $\Zsun$, $0.45\Zsun$ and $0.1\Zsun$.
The binary is detached until Roche lobe overflow starts, as indicated by the black, dashed line. This line is representative of the complete mass transfer phase because the mass transfer occurs on a thermal timescale. 
At all metallicities, the mass transfer does not remove the entire envelope leaving behind around $9\Msun$ of hydrogen-rich material ($9.4\Msun$, $8.5\Msun$ and $8.4\Msun$ for \Zsun, 0.45\Zsun, and 0.1\Zsun, respectively) when the donor star shrinks back into its Roche lobe. We note that \posydon uses an inefficient semi-convection ($\alpha_\mathrm{sc}=0.1$), which leads to more efficient envelope stripping at low $Z$ and easier WR formation compared to a more efficient semi-convection \citep{Klencki+22}.

At $\Zsun$, the remaining hydrogen is quickly removed by strong stellar winds which make the star appear as a WR star quickly after mass transfer has ceased.
At $0.45\Zsun$, removing the hydrogen envelope takes slightly longer due to weaker WR winds. Although the star has crossed the $X_H<0.5$ boundary to be a WR star, the winds are not yet optically thick to be classified as a WR star. This leads to a later start of the WR phase that does not coincide with the mass transfer phase as it does in the $\Zsun$ model. 
At $0.1\Zsun$, a similar sized envelope remains after the mass transfer compared to the higher metallicity models, but the WR winds are not strong enough to remove it before reaching core-carbon depletion. As a result of the weaker WR winds, the optical depth is also insufficiently high for the partially stripped star to appear as a WR star, instead it stops its transition to the hot, luminous side of the Hertzsprung-Russell diagram and spends the remaining part of its life as a partially-stripped star \citep[][see also the HR diagram in Appendix \ref{app:HR_diagram_0.1Zsun}]{Klencki+22}.
While it also has been suggested that Case-B mass transfer might be unstable to explain the absence of long-period WR binaries \citep{Deshmukh+24}, the example models shown in Figure~\ref{fig:example_case_B} show that partial stripping through stable mass transfer and WR wind strength can explain why the contribution from Case-B mass transfer disappears towards lower metallicity.

\subsection{WR formation through Case-A} \label{sec:Case_A}

\begin{figure}
    \centering
    \includegraphics[width=\linewidth]{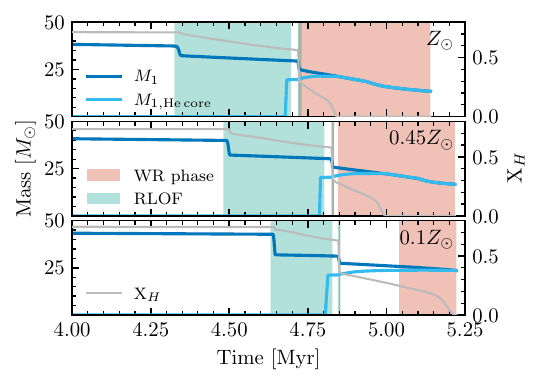}
    \caption{Evolution of an example model undergoing Case-A SMT with $P_\mathrm{ZAMS}=10$ days, $\mathrm{M}_1=45\,\Msun$, and $q=0.8$ at $\Zsun$ (top panel), $0.45\Zsun$ (middle panel) and $0.1\Zsun$ (top panel). The dark and light blue line shows the total primary star mass and the helium core mass, respectively. Roche lobe overflow is indicated as the pink area. All three models undergo a fast Case-A mass transfer, followed by a slow nuclear timescale mass transfer. The systems detaches when nearing the end of the main-sequence and undergo Case-B mass transfer when it leaves the main-sequence. The orange area is when the model falls within the WR criteria, as discussed in Section \ref{sec:methods}. The gray line and right y-axis shows the surface hydrogen abundance.}
    \label{fig:example_case_A}
\end{figure}

Although the contribution of Case-A mass transfer decreases with metallicity, it is not affected in the same way as Case-B mass transfer systems are. 
Figure \ref{fig:example_case_A} shows the same binaries as in Figure \ref{fig:example_case_B} but at $P_\mathrm{ZAMS}=10$ days. All three models undergo mass transfer on the main-sequence with first a phase of fast-mass transfer followed by a slower, nuclear timescale mass transfer due to the reversal of the mass ratio \citep[see also][in the context of Algol binaries]{Sen+22}.
Independent of metallicity, the models detach with around ${\sim}10\Msun$ of hydrogen envelope left.
Due to this left-over hydrogen, the donor star expands again when it reaches core hydrogen exhaustion, initiating Case-B mass transfer. A few more solar masses are stripped from the star, leaving 4.3\Msun, 4.7\Msun, and 5.8\Msun of hydrogen envelope at \Zsun, 0.45\Zsun, and 0.1\Zsun, respectively.
Compared to the Case-B systems, this is half the envelope mass. For \Zsun, this leads to an immediate WR phase, but for 0.45\Zsun and 0.1\Zsun, the winds need to remove some of the rest of the envelope. Even for 0.1\Zsun, the envelope is small enough to be completely stripped before reaching core-carbon depletion and the WR winds strong enough for the star to be classified as a WR star, based on the optical depth criteria. This mechanism is similar to the proposed binary formation scenario for LMC173-1, as proposed by \citet{Cocke+26}, except that in our models the WR phase is only reached after a second phase of mass transfer.
Furthermore, \citet{Nuijten+25} performed an analytical analysis of 21 observed WR+O binaries in the Galaxy to infer the possible values for initial masses, mass transfer efficiency, and angular momentum loss. Although they find a requirement for this mass transfer to be highly non-conservative, their finding of Case-A mass transfer being the most likely scenario of the observed WR+O binaries formation aligns with our assessment of the Case-A peak in the WR period distribution. The combination of multiple mass transfer phases from short ZAMS periods allows the donor to be sufficiently stripped for it to experience a WR phase, despite a leftover hydrogen envelope and weakened WR winds.

Despite their different mass transfer histories, the two \Zsun example binaries in Figure \ref{fig:example_case_B} and \ref{fig:example_case_A} show a similar total orbital widening, expanding by a factor of ${\sim}3$ from ZAMS to core-carbon depletion. Stable Case-A mass transfer can occur down to $P_\mathrm{ZAMS}\sim 3$ days, depending on the component masses. Applying the same widening factor to this shorter initial period yields a final binary period of ${\sim}9$ days. On the other hand, stable Case-B mass transfer requires an initially wider period: with an initial period of ${\sim}400$ days, its final period is ${\sim}1200$ days. Though both systems spends most of its post-mass-transfer lifetime at shorter periods before reaching this final value, the same orbital widening factor in both models leads to a more drastic shift in initially wider systems. At lower metallicities, the initial period at which stable mass transfer occurs decreases further \citep[see][]{Briel+26}, and less mass is lost to stellar winds, allowing even shorter-period WR binaries to form.

\subsection{Comparison against other works}

Several previous population synthesis studies have examined aspects of the WR binary population \citep[e.g.][]{Vanbeveren+98a, Shara+16, Eldridge+17, Stanway+20a, Kasdagli+26}, typically focusing on the ratios between stellar types without explicitly determining the properties of WR binaries. Notable exceptions are the works by \citet{Pauli+22, Pauli+26} and \citet{Xu+25}, who provide period distributions of WR+OB systems using detailed binary models for the LMC and SMC, respectively. These studies consider only the contribution of the primary star to the WR population, whereas we additionally include the contribution from the secondary. These studies find a similar preference for short-period WR binaries as in our models, but also contain an extended tail of long-period WR binaries at LMC and SMC metallicities. By probing a larger metallicity range with the \posydon models, we show that these Case-B mass transfer systems stop contributing as you decrease the metallicity further.

It is important to note that these previous works use an initial period distribution based on \citet{Sana+12}: \citet{Pauli+22} assumes a power-law distribution with a slope of -0.45 by \citet{Sana+13}, while \citet{Xu+25} assumes the same distribution with -0.55 \citep{Sana+12}. These negative slopes on the period distribution highly favor the formation of short initial periods and suppresses long-period systems. 
In our work, we have assumed a flat $\log(P/\mathrm{days})$ distribution, which is equivalent to a power-law slope of 0, more closely resembling more recent observational results \citep{Sana+25}. Due to this flatter period distribution, the \posydon models have a larger Case-B contribution than the other studies at the same metallicity (For the \posydon populations with the \citet{Sana+12} initial period distribution, see Appendix \ref{app:initial_period_dist}).

In addition to the differences in initial period distribution, Case-B mass transfer is treated differently. The binary models from \citet{Pauli+22} and \citet{Xu+25} both use the \texttt{contact} scheme throughout their evolution, which limits the donor star to the Roche lobe, while \posydon switches to the \texttt{kolb} scheme after the main-sequence to allow for radial expansion past the Roche lobe. This can lead to additional differences in the contribution of Case-B systems and its metallicity dependence. While \citet{Pauli+22} uses the \citet{Nugis+00} WR mass loss rates, similar to the \posydon models, \citet{Xu+25} instead uses WR wind mass loss rates of \citet{Hamann+95}. This can results in differences in the contribution of post-Case-B MT systems, as the amount of wind mass loss determines if these systems can contribute to the WR population. Beyond these studies of WR+OB systems, \citet{Langer+20} has made predictions for the LMC for the population of OB+BH systems under the assumption that the WR star collapses to form a black hole, which exhibit the same Case-A and Case-B peaks as in our WR binary population.

Taken together, while our default models show a stronger Case-B peak than \citet{Pauli+22} and \citet{Xu+25} at the same metallicities due to our flatter initial period distribution, the broader range of metallicities explored in this work provides an explanation for the observed absence of wide-period WR binaries that is not accessible within the more limited metallicity coverage of previous studies.






\section{Conclusions} \label{sec:conclusion}

Using detailed binary models, we have predicted the period distribution of WR binaries at $0.01\Zsun$, $0.1\Zsun$, $0.2\Zsun$, $0.45\Zsun$, and $\Zsun$ and investigated their formation mechanism. We include both the primary and secondary in binaries producing WRs, where the latter adds an additional contribution of ${\sim}30\%$. Overall, we find that SMT and non-interacting systems are the dominant formation scenario of WRs in binaries. Moreover, due to a combination of partial stripping and decreased WR stellar wind strength at low metallicity, we find an absence of long-period WR binaries at low metallicity, providing a possible explanation for the absence of long-period systems in SMC and LMC observations \citep{Shenar+16, Shenar+19, Schootemeijer+24}. This results appears because of the interplay between the mass transfer and stellar wind strength across across metallicity that is only resolvable with detailed binary models.
We additionally find the same relative contributions to the WR binary population from the non-interacting and SMT channel across the explored metallicities, implying that a constant WR binary fraction is expected across metallicity.

Furthermore, we find that the SMT channel produces two peaks in the WR period distribution: a short-period peak from Case-A mass transfer, and a long-period peak from Case-B mass transfer. The Case-A WR binary population peaks at $\sim20$ days at $\Zsun$ which shift to ${\sim}4$ days at $0.1\Zsun$, aligning closely to the peak in the observed WR binaries in the SMC, LMC, and Galactic populations. The Case-B mass transfer peak, on the other hand, produces a peak at longer periods (500 days), similar to the peaks found by \citet{Langer+20} for O/B+black hole type systems. However, we find that this peak disappears below $0.2\Zsun$ as WR stellar winds are no longer strong enough to remove the remaining hydrogen envelope after mass transfer, which leads to the absence of long-period WR binaries at low metallicity.

We additionally find that common envelope evolution only produces a small number of WR binaries in \posydon, but we have not explored the effect of the CE parameter in this work. Most CE with a stellar companion lead to a merger. Only at sub-solar metallicity do more systems with a stellar companion survive the CE but their number is still limited.
On the other hand, at $\Zsun$, CEE with a BH companion produce tight-orbit WR systems with $P<1$ day, which cannot be produced by any other formation mechanism in \posydon.
If this formation mechanism is robust, WRs in tight-orbit binaries, such as Cygnus X-3, NGC 300 X-1, and IC 10 X-1, are very likely post-CE systems, providing a clear testbed for common envelope physics \citep[for an individual example, see][]{Wong+14}.

Crucially, the different formation channels of WR binaries are sensitive to different stellar winds. We find that long-period non-interacting WR binaries, especially at solar metallicity, are sensitive to main-sequence and RSG winds, while another sub-population of non-interacting WR at a range of orbital periods is sensitive to the main-sequence and LBV wind strengths. The presence of the Case-B mass transfer peak, on the other hand, depends on the strength of the WR wind mass loss. Since each of these leaves an imprint on the period distribution of WR binaries, these features can be used as benchmarks for stellar winds.

However, although the WR binary populations presented here provide a possible answer to the possible observed absence of long-period WR binaries in observations, the synthetic population from \posydon do not align perfectly with the observations. For example the presence of long-period WRs from Case-B mass transfer at SMC-like and LMC-like metallicities. These discrepancies are either a stellar wind modeling effect or an observational bias, as the SMC only contains a small number of WR binaries. While the statistics of the LMC and Galactic WR population are higher, they have not had the same systematic searches with modern surveys for WR binaries yet. Future observations will hopefully lead to a more complete picture of WR binaries and their properties, allowing us to use these populations as strong constraints for stellar and binary modeling.

Finding specific stellar population in which the imprints of single-star and binary-interaction assumptions can be separated, provides a crucial testing ground for both single-star evolution and binary effects in binary population synthesis. Correctly reproducing these populations is a prerequisite for accurately modeling their descendants, such as stripped-envelope supernovae and compact object mergers.

\begin{acknowledgements}

The authors thank Andreas Sander, Kunal Deshmukh, and Dori (Dorottya) Szécsi for the helpful discussions and questions. We thank David Aguilera-Dena for the insightful comments on the manuscript. The \posydon project is supported primarily by two sources: the Swiss National Science Foundation (PI Fragos, project number CRSII5\_213497) and the Gordon and Betty Moore Foundation (PI Kalogera, grant award GBMF8477 and GBMF12341). E.K.\ and J.J.A.\ acknowledge support for Program number (JWST-AR-04369.001-A) provided through a grant from the STScI under NASA contract NAS5-03127. The computations were performed at Northwestern University on the Trident computer cluster (funded by the GBMF8477 award) and at the University of Geneva on the Yggdrasil computer cluster. This research was supported in part through the computational resources and staff contributions provided for the Quest high performance computing facility at Northwestern University which is jointly supported by the Office of the Provost, the Office for Research, and Northwestern University Information Technology. TS acknowledges support from the European Research Council (ERC) under the European Union's Horizon 2020 research and innovation programme (grant agreement 101164755/METAL) and the Israel Science Foundation (ISF) under grant number 0603225041. S.G.\ is supported by the Gordon and Betty Moore Foundation (grant award GBMF12341).
\end{acknowledgements}

%

\bibliographystyle{aa}
\bibliography{WR_binaries} 

@article{Abbott+87,
  title = {Wolf-Rayet Stars.},
  author = {Abbott, David C. and Conti, Peter S.},
  year = 1987,
  month = jan,
  journal = {ARA\&A},
  volume = {25},
  pages = {113--150},
  issn = {0066-4146},
  doi = {10.1146/annurev.aa.25.090187.000553},
  urldate = {2026-07-03}
}

@article{Aguilera-Dena+22,
  title = {Stripped-Envelope Stars in Different Metallicity Environments. {{I}}. {{Evolutionary}} Phases, Classification, and Populations},
  author = {{Aguilera-Dena}, David R. and Langer, Norbert and Antoniadis, John and Pauli, Daniel and Dessart, Luc and {Vigna-G{\'o}mez}, Alejandro and Gr{\"a}fener, G{\"o}tz and Yoon, Sung-Chul},
  year = 2022,
  month = may,
  journal = {A\&A},
  volume = {661},
  pages = {A60},
  issn = {0004-6361},
  doi = {10.1051/0004-6361/202142895},
  urldate = {2024-10-31}
}

@article{Andrews+25,
  title = {{{POSYDON}} Version 2: Population Synthesis with Detailed Binary-Evolution Simulations across a Cosmological Range of Metallicities},
  shorttitle = {{{POSYDON}} Version 2},
  author = {Andrews, Jeff J. and Bavera, Simone S. and Briel, Max and Chattaraj, Abhishek and Dotter, Aaron and Fragos, Tassos and {Gallegos-Garcia}, Monica and Gossage, Seth and Kalogera, Vicky and Kasdagli, Eirini and Katsaggelos, Aggelos and Kimball, Chase and Kovlakas, Konstantinos and Kruckow, Matthias U. and Liotine, Camille and Misra, Devina and Rocha, Kyle A. and Souropanis, Dimitris and Srivastava, Philipp M. and Sun, Meng and Teng, Elizabeth and Xing, Zepei and Zapartas, Emmanouil and Zevin, Michael},
  year = 2025,
  month = nov,
  journal = {ApJS},
  volume = {281},
  pages = {3--51},
  publisher = {IOP},
  issn = {0067-0049},
  doi = {10.3847/1538-4365/adfb78},
  urldate = {2026-01-14}
}

@article{Antoniadis+25,
  title = {Investigating the Metallicity Dependence of the Mass-Loss Rate Relation of Red Supergiants},
  author = {Antoniadis, K. and Zapartas, E. and Bonanos, A. Z. and Maravelias, G. and Vlassis, S. and {Mu{\~n}oz-Sanchez}, G. and Nally, C. and Meixner, M. and Jones, O. C. and Lenki{\'c}, L. and Kavanagh, P. J.},
  year = 2025,
  month = oct,
  journal = {A\&A},
  volume = {702},
  pages = {A178},
  publisher = {EDP},
  issn = {0004-6361},
  doi = {10.1051/0004-6361/202554416},
  urldate = {2026-08-12}
}

@article{Bartzakos+01,
  title = {Magellanic {{Cloud WC}}/{{WO Wolf-Rayet}} Stars - {{I}}. {{Binary}} Frequency and {{Roche}} Lobe Overflow Formation},
  author = {Bartzakos, P. and Moffat, A. F. J. and Niemela, V. S.},
  year = 2001,
  month = jun,
  journal = {MNRAS},
  volume = {324},
  pages = {18--32},
  publisher = {OUP},
  issn = {0035-8711},
  doi = {10.1046/j.1365-8711.2001.04126.x},
  urldate = {2026-07-03}
}

@article{Belczynski+10a,
  title = {On the {{Maximum Mass}} of {{Stellar Black Holes}}},
  author = {Belczynski, Krzysztof and Bulik, Tomasz and Fryer, Chris L. and Ruiter, Ashley and Valsecchi, Francesca and Vink, Jorick S. and Hurley, Jarrod R.},
  year = 2010,
  month = may,
  journal = {ApJ},
  volume = {714},
  pages = {1217--1226},
  publisher = {IOP},
  issn = {0004-637X},
  doi = {10.1088/0004-637X/714/2/1217},
  urldate = {2024-10-31}
}

@article{Belczynski+13,
  title = {Cyg {{X-3}}: A Galactic Double Black Hole or Black Hole-Neutron Star Progenitor},
  shorttitle = {Cyg {{X-3}}},
  author = {Belczynski, Krzysztof and Bulik, Tomasz and Mandel, Ilya and Sathyaprakash, B. S. and Zdziarski, Andrzej and Mikolajewska, Joanna},
  year = 2013,
  month = jan,
  journal = {ApJ},
  volume = {764},
  number = {1},
  eprint = {1209.2658},
  primaryclass = {astro-ph.HE},
  pages = {96},
  issn = {0004-637X, 1538-4357},
  doi = {10.1088/0004-637X/764/1/96},
  urldate = {2026-07-22},
  archiveprefix = {arXiv}
}

@article{Bhargava+17,
  title = {A Precise Measurement of the Orbital Period Parameters of Cygnus {{X-3}}},
  author = {Bhargava, Yash and Rao, A. R. and Singh, K. P. and Choudhury, Manojendu and Bhattacharyya, S. and Chandra, S. and Dewangan, G. C. and Mukerjee, K. and Stewart, G. C. and Bhattacharya, D. and Mithun, N. P. S. and Vadawale, S. V.},
  year = 2017,
  month = nov,
  journal = {ApJ},
  volume = {849},
  number = {2},
  eprint = {1709.07441},
  primaryclass = {astro-ph.HE},
  pages = {141},
  issn = {0004-637X, 1538-4357},
  doi = {10.3847/1538-4357/aa8ea4},
  urldate = {2026-07-10},
  archiveprefix = {arXiv}
}

@article{Binder+21,
  title = {The Wolf-Rayet + Black Hole Binary {{NGC}} 300 {{X-1}}: What Is the Mass of the Black Hole?},
  shorttitle = {The Wolf-Rayet + Black Hole Binary {{NGC}} 300 {{X-1}}},
  author = {Binder, Breanna A. and Sy, Janelle M. and Eracleous, Michael and Christodoulou, Dimitris M. and Bhattacharya, Sayantan and Cappallo, Rigel and Laycock, Silas and Plucinsky, Paul P. and Williams, Benjamin F.},
  year = 2021,
  month = mar,
  journal = {ApJ},
  volume = {910},
  number = {1},
  eprint = {2102.07065},
  primaryclass = {astro-ph},
  pages = {74},
  issn = {0004-637X, 1538-4357},
  doi = {10.3847/1538-4357/abe6a9},
  urldate = {2026-03-09},
  archiveprefix = {arXiv}
}

@article{Bloecker+95,
  title = {Stellar Evolution of Low and Intermediate-Mass Stars. {{I}}. {{Mass}} Loss on the {{AGB}} and Its Consequences for Stellar Evolution},
  author = {Bloecker, T.},
  year = 1995,
  month = may,
  journal = {A\&A},
  volume = {297},
  pages = {727},
  publisher = {EDP},
  issn = {0004-6361},
  urldate = {2026-07-03}
}

@article{Briel+26,
  title = {A Case for Case a: Detailed Look at Binary Black Hole Formation through Stable Mass Transfer},
  shorttitle = {A Case for Case a},
  author = {Briel, Max M. and Fragos, Tassos and {Gallegos-Garcia}, Monica and Ray, Anarya and Zevin, Michael and Chattaraj, Abhishek and Andrews, Jeff J. and Kalogera, Vicky and Gossage, Seth and Srivastava, Philipp M. and Teng, Elizabeth},
  year = 2026,
  month = feb,
  publisher = {arXiv},
  doi = {10.48550/arXiv.2602.03629},
  urldate = {2026-04-15}
}

@article{Cocke+26,
  title = {An Orbit for a Massive Wolf-Rayet Binary in the {{LMC}}: An Example of Binary Evolution},
  shorttitle = {An Orbit for a Massive Wolf-Rayet Binary in the {{LMC}}},
  author = {Cocke, Breelyn and Massey, Philip and Morrell, Nidia I. and Penny, Laura R. and Neugent, Kathryn F. and Eldridge, Jan J. and Szymanski, Michal K. and Udalski, Andrzej and Marin, Laurella C.},
  year = 2026,
  month = mar,
  number = {arXiv:2603.11357},
  eprint = {2603.11357},
  primaryclass = {astro-ph},
  publisher = {arXiv},
  doi = {10.48550/arXiv.2603.11357},
  urldate = {2026-03-13},
  archiveprefix = {arXiv}
}

@article{Conti+75,
  title = {On the Relationship between {{Of}} and {{WR}} Stars.},
  author = {Conti, P. S.},
  year = 1975,
  month = jan,
  journal = {Memoires Of Societe Royale Des Sciences De Liege},
  volume = {9},
  pages = {193--212},
  urldate = {2024-10-31}
}

@article{Crowther+07,
  title = {Physical {{Properties}} of {{Wolf-Rayet Stars}}},
  author = {Crowther, Paul A.},
  year = 2007,
  month = sep,
  journal = {ARA\&A},
  volume = {45},
  pages = {177--219},
  issn = {0066-4146},
  doi = {10.1146/annurev.astro.45.051806.110615},
  urldate = {2026-07-03}
}

@article{Crowther+10,
  title = {{{NGC}} 300 {{X-1}} Is a Wolf-Rayet/Black Hole Binary},
  author = {Crowther, P. A. and Barnard, R. and Carpano, S. and Clark, J. S. and Dhillon, V. S. and Pollock, A. M. T.},
  year = 2010,
  month = mar,
  journal = {MNRAS},
  volume = {403},
  pages = {L41-L45},
  publisher = {OUP},
  issn = {0035-8711},
  doi = {10.1111/j.1745-3933.2010.00811.x},
  urldate = {2026-03-09}
}

@article{deJager+88,
  title = {Mass Loss Rates in the {{Hertzsprung-Russell}} Diagram.},
  author = {{de Jager}, C. and Nieuwenhuijzen, H. and {van der Hucht}, K. A.},
  year = 1988,
  month = feb,
  journal = {A\&A, Suppl. Ser.},
  volume = {72},
  pages = {259--289},
  publisher = {EDP},
  issn = {0365-0138},
  urldate = {2026-07-03}
}

@article{deKoter+97,
  title = {On the Evolutionary Phase and Mass Loss of the Wolf-Rayet-like Stars in {{R136a}}*},
  author = {{de Koter}, Alex and Heap, Sara R. and Hubeny, Ivan},
  year = 1997,
  month = mar,
  journal = {ApJ},
  volume = {477},
  number = {2},
  pages = {792},
  publisher = {IOP Publishing},
  issn = {0004-637X},
  doi = {10.1086/303736},
  urldate = {2026-07-03}
}

@article{deMink+13,
  title = {The Rotation Rates of Massive Stars: The Role of Binary Interaction through Tides, Mass Transfer, and Mergers},
  shorttitle = {The Rotation Rates of Massive Stars},
  author = {{de Mink}, S. E. and Langer, N. and Izzard, R. G. and Sana, H. and {de Koter}, A.},
  year = 2013,
  month = feb,
  journal = {ApJ},
  volume = {764},
  pages = {166},
  issn = {0004-637X},
  doi = {10.1088/0004-637X/764/2/166},
  urldate = {2023-05-10}
}

@article{Deshmukh+24,
  title = {Investigating 39 Galactic Wolf-Rayet Stars with {{VLTI}}/{{GRAVITY}}: Uncovering a Long-Period Binary Desert},
  shorttitle = {Investigating 39 Galactic Wolf-Rayet Stars with {{VLTI}}/{{GRAVITY}}},
  author = {Deshmukh, K. and Sana, H. and M{\'e}rand, A. and Bordier, E. and Langer, N. and Bodensteiner, J. and Dsilva, K. and Frost, A. J. and Gosset, E. and Le Bouquin, J.-B. and Lefever, R. R. and Mahy, L. and Patrick, L. R. and Reggiani, M. and Sander, A. A. C. and Shenar, T. and Tramper, F. and Villase{\~n}or, J. I. and Waisberg, I.},
  year = 2024,
  month = dec,
  journal = {A\&A},
  volume = {692},
  pages = {A109},
  publisher = {EDP},
  issn = {0004-6361},
  doi = {10.1051/0004-6361/202452352},
  urldate = {2025-12-15}
}

@article{Drout+23,
  title = {An Observed Population of Intermediate-Mass Helium Stars That Have Been Stripped in Binaries},
  author = {Drout, M. R. and G{\"o}tberg, Y. and Ludwig, B. A. and Groh, J. H. and {de Mink}, S. E. and O'Grady, A. J. G. and Smith, N.},
  year = 2023,
  month = dec,
  journal = {Science},
  volume = {382},
  pages = {1287--1291},
  issn = {0036-8075},
  doi = {10.1126/science.ade4970},
  urldate = {2024-10-31}
}

@article{Dsilva+20,
  title = {A Spectroscopic Multiplicity Survey of {{Galactic Wolf-Rayet}} Stars. {{I}}. {{The}} Northern {{WC}} Sequence},
  author = {Dsilva, K. and Shenar, T. and Sana, H. and Marchant, P.},
  year = 2020,
  month = sep,
  journal = {A\&A},
  volume = {641},
  pages = {A26},
  issn = {0004-6361},
  doi = {10.1051/0004-6361/202038446},
  urldate = {2024-10-31}
}

@article{Dsilva+22,
  title = {A Spectroscopic Multiplicity Survey of {{Galactic Wolf-Rayet}} Stars. {{II}}. {{The}} Northern {{WNE}} Sequence},
  author = {Dsilva, K. and Shenar, T. and Sana, H. and Marchant, P.},
  year = 2022,
  month = aug,
  journal = {A\&A},
  volume = {664},
  pages = {A93},
  issn = {0004-6361},
  doi = {10.1051/0004-6361/202142729},
  urldate = {2024-10-31}
}

@article{Dsilva+23,
  title = {A Spectroscopic Multiplicity Survey of {{Galactic Wolf-Rayet}} Stars . {{III}}. {{The}} Northern Late-Type Nitrogen-Rich Sample},
  author = {Dsilva, K. and Shenar, T. and Sana, H. and Marchant, P.},
  year = 2023,
  month = jun,
  journal = {A\&A},
  volume = {674},
  pages = {A88},
  issn = {0004-6361},
  doi = {10.1051/0004-6361/202244308},
  urldate = {2024-10-31}
}

@article{Eldridge+17,
  title = {Binary {{Population}} and {{Spectral Synthesis Version}} 2.1: {{Construction}}, {{Observational Verification}}, and {{New Results}}},
  shorttitle = {Binary {{Population}} and {{Spectral Synthesis Version}} 2.1},
  author = {Eldridge, J. J. and Stanway, E. R. and Xiao, L. and McClelland, L. A. S. and Taylor, G. and Ng, M. and Greis, S. M. L. and Bray, J. C.},
  year = 2017,
  month = nov,
  journal = {PASA},
  volume = {34},
  eprint = {1710.02154v1},
  pages = {e058},
  issn = {1323-3580},
  doi = {10.1017/pasa.2017.51},
  urldate = {2022-05-25},
  archiveprefix = {arXiv}
}

@article{Foellmi+03,
  title = {Wolf-Rayet Binaries in the Magellanic Clouds and Implications for Massive-Star Evolution - {{II}}. {{Large}} Magellanic Cloud},
  author = {Foellmi, C. and Moffat, A. F. J. and Guerrero, M. A.},
  year = 2003,
  month = feb,
  journal = {MNRAS},
  volume = {338},
  pages = {1025--1056},
  publisher = {OUP},
  issn = {0035-8711},
  doi = {10.1046/j.1365-8711.2003.06161.x},
  urldate = {2026-02-23}
}

@article{Foellmi+03a,
  title = {Wolf-Rayet Binaries in the Magellanic Clouds and Implications for Massive-Star Evolution - {{I}}. {{Small}} Magellanic Cloud},
  author = {Foellmi, C. and Moffat, A. F. J. and Guerrero, M. A.},
  year = 2003,
  month = jan,
  journal = {MNRAS},
  volume = {338},
  pages = {360--388},
  publisher = {OUP},
  issn = {0035-8711},
  doi = {10.1046/j.1365-8711.2003.06052.x},
  urldate = {2026-02-23}
}

@article{Fragos+19,
  title = {The Complete Evolution of a Neutron-Star Binary through a Common Envelope Phase Using {{1D}} Hydrodynamic Simulations},
  author = {Fragos, Tassos and Andrews, Jeff J. and {Ramirez-Ruiz}, Enrico and Meynet, Georges and Kalogera, Vicky and Taam, Ronald E. and Zezas, Andreas},
  year = 2019,
  month = oct,
  journal = {ApJ},
  volume = {883},
  pages = {L45},
  publisher = {IOP},
  issn = {0004-637X},
  doi = {10.3847/2041-8213/ab40d1},
  urldate = {2026-04-23}
}

@article{Fragos+23,
  title = {{{POSYDON}}: A General-Purpose Population Synthesis Code with Detailed Binary-Evolution Simulations},
  shorttitle = {Posydon},
  author = {Fragos, Tassos and Andrews, Jeff J. and Bavera, Simone S. and Berry, Christopher P. L. and Coughlin, Scott and Dotter, Aaron and Giri, Prabin and Kalogera, Vicky and Katsaggelos, Aggelos and Kovlakas, Konstantinos and Lalvani, Shamal and Misra, Devina and Srivastava, Philipp M. and Qin, Ying and Rocha, Kyle A. and {Roman-Garza}, Jaime and Serra, Juan Gabriel and Stahle, Petter and Sun, Meng and Teng, Xu and Trajcevski, Goce and Tran, Nam Hai and Xing, Zepei and Zapartas, Emmanouil and Zevin, Michael},
  year = 2023,
  month = feb,
  journal = {ApJS},
  volume = {264},
  number = {2},
  eprint = {2202.05892},
  primaryclass = {astro-ph},
  pages = {45--105},
  issn = {0067-0049, 1538-4365},
  doi = {10.3847/1538-4365/ac90c1},
  urldate = {2024-01-26},
  archiveprefix = {arXiv}
}

@article{Fryer+12,
  title = {Compact Remnant Mass Function: Dependence on the Explosion Mechanism and Metallicity},
  shorttitle = {Compact Remnant Mass Function},
  author = {Fryer, Chris L. and Belczynski, Krzysztof and Wiktorowicz, Grzegorz and Dominik, Michal and Kalogera, Vicky and Holz, Daniel E.},
  year = 2012,
  month = mar,
  journal = {ApJ},
  volume = {749},
  number = {1},
  pages = {91},
  publisher = {American Astronomical Society},
  issn = {0004-637X},
  doi = {10.1088/0004-637X/749/1/91},
  urldate = {2022-01-07}
}

@article{Ghosh+81,
  title = {The Asymmetric 4.8 Hour {{X-ray}} Modulation of {{CYG X-3}} : Model Light Curves and Inferred Orbital Parameters.},
  shorttitle = {The Asymmetric 4.8 Hour {{X-ray}} Modulation of {{CYG X-3}}},
  author = {Ghosh, P. and Elsner, R. F. and Weisskopf, M. C. and Sutherland, P. G.},
  year = 1981,
  month = dec,
  journal = {ApJ},
  volume = {251},
  pages = {230--245},
  publisher = {IOP},
  issn = {0004-637X},
  doi = {10.1086/159458},
  urldate = {2026-07-10}
}

@article{Gotberg+17,
  title = {Ionizing Spectra of Stars That Lose Their Envelope through Interaction with a Binary Companion: Role of Metallicity},
  shorttitle = {Ionizing Spectra of Stars That Lose Their Envelope through Interaction with a Binary Companion},
  author = {G{\"o}tberg, Y. and de Mink, S. E. and Groh, J. H.},
  year = 2017,
  month = dec,
  journal = {A\&A},
  volume = {608},
  pages = {A11},
  publisher = {EDP Sciences},
  issn = {0004-6361, 1432-0746},
  doi = {10.1051/0004-6361/201730472},
  urldate = {2022-03-31},
  copyright = {\copyright{} ESO, 2017}
}

@article{Groh+14,
  title = {The Evolution of Massive Stars and Their Spectra. {{I}}. {{A}} Non-Rotating 60 {{M}}{$\Sun$} Star from the Zero-Age Main Sequence to the Pre-Supernova Stage},
  author = {Groh, Jose H. and Meynet, Georges and Ekstr{\"o}m, Sylvia and Georgy, Cyril},
  year = 2014,
  month = apr,
  journal = {A\&A},
  volume = {564},
  pages = {A30},
  publisher = {EDP},
  issn = {0004-6361},
  doi = {10.1051/0004-6361/201322573},
  urldate = {2026-08-12}
}

@article{Hainich+15,
  title = {Wolf-Rayet Stars in the Small Magellanic Cloud. {{I}}. {{Analysis}} of the Single {{WN}} Stars},
  author = {Hainich, R. and Pasemann, D. and Todt, H. and Shenar, T. and Sander, A. and Hamann, W.-R.},
  year = 2015,
  month = sep,
  journal = {A\&A},
  volume = {581},
  pages = {A21},
  publisher = {EDP},
  issn = {0004-6361},
  doi = {10.1051/0004-6361/201526241},
  urldate = {2026-02-23}
}

@article{Hamann+06,
  title = {The {{Galactic WN}} Stars. {{Spectral}} Analyses with Line-Blanketed Model Atmospheres versus Stellar Evolution Models with and without Rotation},
  author = {Hamann, W.-R. and Gr{\"a}fener, G. and Liermann, A.},
  year = 2006,
  month = oct,
  journal = {A\&A},
  volume = {457},
  pages = {1015--1031},
  publisher = {EDP},
  issn = {0004-6361},
  doi = {10.1051/0004-6361:20065052},
  urldate = {2026-07-03}
}

@article{Hamann+95,
  title = {Spectral Analyses of the {{Galactic Wolf-Rayet}} Stars: Hydrogen-Helium Abundances and Improved Stellar Parameters for the {{WN}} Class},
  shorttitle = {Spectral Analyses of the {{Galactic Wolf-Rayet}} Stars},
  author = {Hamann, W.-R. and Koesterke, L. and Wessolowski, U.},
  year = 1995,
  month = jul,
  journal = {A\&A},
  volume = {299},
  pages = {151},
  publisher = {EDP},
  issn = {0004-6361},
  urldate = {2026-07-10}
}

@article{Humphreys+79,
  title = {Studies of Luminous Stars in Nearby Galaxies. {{III}}. {{Comments}} on the Evolution of the Most Massive Stars in the {{Milky Way}} and the {{Large Magellanic Cloud}}.},
  author = {Humphreys, R. M. and Davidson, K.},
  year = 1979,
  month = sep,
  journal = {ApJ},
  volume = {232},
  pages = {409--420},
  publisher = {IOP},
  issn = {0004-637X},
  doi = {10.1086/157301},
  urldate = {2026-07-03}
}

@article{Ivanova+13,
  title = {Common Envelope Evolution: Where We Stand and How We Can Move Forward},
  shorttitle = {Common Envelope Evolution},
  author = {Ivanova, N. and Justham, S. and Chen, X. and De Marco, O. and Fryer, C. L. and Gaburov, E. and Ge, H. and Glebbeek, E. and Han, Z. and Li, X.-D. and Lu, G. and Marsh, T. and Podsiadlowski, {\relax Ph}. and Potter, A. and Soker, N. and Taam, R. and Tauris, T. M. and {van den Heuvel}, E. P. J. and Webbink, R. F.},
  year = 2013,
  month = nov,
  journal = {A\&AR},
  volume = {21},
  number = {1},
  eprint = {1209.4302},
  pages = {59},
  issn = {0935-4956},
  doi = {10.1007/s00159-013-0059-2},
  urldate = {2016-05-10},
  archiveprefix = {arXiv}
}

@article{Jermyn+23,
  title = {Modules for Experiments in Stellar Astrophysics ({{MESA}}): Time-Dependent Convection, Energy Conservation, Automatic Differentiation, and Infrastructure},
  shorttitle = {Modules for Experiments in Stellar Astrophysics ({{MESA}})},
  author = {Jermyn, Adam S. and Bauer, Evan B. and Schwab, Josiah and Farmer, R. and Ball, Warrick H. and Bellinger, Earl P. and Dotter, Aaron and Joyce, Meridith and Marchant, Pablo and Mombarg, Joey S. G. and Wolf, William M. and Sunny Wong, Tin Long and Cinquegrana, Giulia C. and Farrell, Eoin and Smolec, R. and Thoul, Anne and Cantiello, Matteo and Herwig, Falk and Toloza, Odette and Bildsten, Lars and Townsend, Richard H. D. and Timmes, F. X.},
  year = 2023,
  month = mar,
  journal = {ApJS},
  volume = {265},
  pages = {15},
  issn = {0067-0049},
  doi = {10.3847/1538-4365/acae8d},
  urldate = {2023-07-17}
}

@article{Kasdagli+26,
  title = {Stellar {{Population Spectra Incorporating Detailed Binary Evolution}} Using {{POSYDON}}},
  author = {Kasdagli, Eirini and Andrews, Jeff J. and Lehmer, Bret and Townsend, Rich and Zapartas, Manos and Zezas, Andreas and Briel, Max and Fragos, Tassos and Gossage, Seth and Srivastava, Philipp M. and Teng, Elizabeth},
  year = 2026,
  month = jun,
  number = {arXiv:2606.13351},
  eprint = {2606.13351},
  primaryclass = {astro-ph.SR},
  publisher = {arXiv},
  doi = {10.48550/arXiv.2606.13351},
  urldate = {2026-06-23},
  archiveprefix = {arXiv}
}

@article{Klencki+20,
  title = {Massive Donors in Interacting Binaries: Effect of Metallicity},
  shorttitle = {Massive Donors in Interacting Binaries},
  author = {Klencki, Jakub and Nelemans, Gijs and Istrate, Alina G. and Pols, Onno},
  year = 2020,
  month = jun,
  journal = {A\&A},
  volume = {638},
  pages = {A55},
  issn = {0004-6361, 1432-0746},
  doi = {10.1051/0004-6361/202037694},
  urldate = {2022-03-29}
}

@article{Klencki+22,
  title = {Partial-Envelope Stripping and Nuclear-Timescale Mass Transfer from Evolved Supergiants at Low Metallicity},
  author = {Klencki, Jakub and Istrate, Alina and Nelemans, Gijs and Pols, Onno},
  year = 2022,
  month = jun,
  journal = {A\&A},
  volume = {662},
  pages = {A56},
  issn = {0004-6361},
  doi = {10.1051/0004-6361/202142701},
  urldate = {2022-09-19}
}

@article{Kolb+90,
  title = {A Comparative Study of the Evolution of a Close Binary Using a Standard and an Improved Technique for Computing Mass Transfer},
  author = {Kolb, U. and Ritter, H.},
  year = 1990,
  month = sep,
  journal = {A\&A},
  volume = {236},
  pages = {385--392},
  issn = {0004-6361},
  urldate = {2023-05-18}
}

@article{Kroupa+01,
  title = {On the Variation of the Initial Mass Function},
  author = {Kroupa, P.},
  year = 2001,
  month = apr,
  journal = {MNRAS},
  volume = {322},
  number = {2},
  pages = {231--246},
  issn = {0035-8711, 1365-2966},
  doi = {10.1046/j.1365-8711.2001.04022.x},
  urldate = {2021-01-18}
}

@article{Kruckow+24,
  title = {The Formation of Black Holes in Non-Interacting Isolated Binaries: {{Gaia}} Black Holes as Calibrators of Stellar Winds from Massive Stars},
  shorttitle = {The Formation of Black Holes in Non-Interacting Isolated Binaries},
  author = {Kruckow, Matthias U. and Andrews, Jeff J. and Fragos, Tassos and Holl, Berry and Bavera, Simone S. and Briel, Max and Gossage, Seth and Kovlakas, Konstantinos and Rocha, Kyle A. and Sun, Meng and Srivastava, Philipp M. and Xing, Zepei and Zapartas, Emmanouil},
  year = 2024,
  month = dec,
  journal = {A\&A},
  volume = {692},
  pages = {A141},
  publisher = {EDP},
  issn = {0004-6361},
  doi = {10.1051/0004-6361/202452356},
  urldate = {2025-05-19}
}

@article{Langer+20,
  title = {Properties of {{OB}} Star-Black Hole Systems Derived from Detailed Binary Evolution Models},
  author = {Langer, N. and Sch{\"u}rmann, C. and Stoll, K. and Marchant, P. and Lennon, D. J. and Mahy, L. and {de Mink}, S. E. and Quast, M. and Riedel, W. and Sana, H. and Schneider, P. and Schootemeijer, A. and Wang, C. and Almeida, L. A. and Bestenlehner, J. M. and Bodensteiner, J. and Castro, N. and Clark, S. and Crowther, P. A. and Dufton, P. and Evans, C. J. and Fossati, L. and Gr{\"a}fener, G. and Grassitelli, L. and Grin, N. and Hastings, B. and Herrero, A. and {de Koter}, A. and Menon, A. and Patrick, L. and Puls, J. and Renzo, M. and Sander, A. A. C. and Schneider, F. R. N. and Sen, K. and Shenar, T. and {Sim{\'o}n-D{\'i}as}, S. and Tauris, T. M. and Tramper, F. and Vink, J. S. and Xu, X. -T.},
  year = 2020,
  month = jun,
  journal = {A\&A},
  volume = {638},
  pages = {A39},
  issn = {0004-6361},
  doi = {10.1051/0004-6361/201937375},
  urldate = {2025-05-19}
}

@article{Langer+89,
  title = {Standard Models of {{Wolf-Rayet}} Stars.},
  author = {Langer, N.},
  year = 1989,
  month = feb,
  journal = {A\&A},
  volume = {210},
  pages = {93--113},
  publisher = {EDP},
  issn = {0004-6361},
  urldate = {2026-01-20}
}

@article{Livio+88,
  title = {The {{Common Envelope Phase}} in the {{Evolution}} of {{Binary Stars}}},
  author = {Livio, Mario and Soker, Noam},
  year = 1988,
  month = jun,
  journal = {ApJ},
  volume = {329},
  pages = {764},
  publisher = {IOP},
  issn = {0004-637X},
  doi = {10.1086/166419},
  urldate = {2026-08-11}
}

@article{Marchant+16,
  title = {A New Route towards Merging Massive Black Holes},
  author = {Marchant, Pablo and Langer, Norbert and Podsiadlowski, Philipp and Tauris, Thomas M. and Moriya, Takashi J.},
  year = 2016,
  month = apr,
  journal = {A\&A},
  volume = {588},
  pages = {A50},
  issn = {0004-6361, 1432-0746},
  doi = {10.1051/0004-6361/201628133},
  urldate = {2024-09-23}
}

@article{Massey+81,
  title = {The Masses of Wolf-Rayet Stars},
  author = {Massey, P.},
  year = 1981,
  month = may,
  journal = {ApJ},
  volume = {246},
  pages = {153--160},
  publisher = {IOP},
  issn = {0004-637X},
  doi = {10.1086/158908},
  urldate = {2026-02-23}
}

@article{Moffat+86,
  title = {Photometric Variability of a Complete Sample of Northern {{Wolf-Rayet}} Stars.},
  author = {Moffat, A. F. J. and Shara, M. M.},
  year = 1986,
  month = oct,
  journal = {Astronomical Journal},
  volume = {92},
  pages = {952--975},
  publisher = {IOP},
  issn = {0004-6256},
  doi = {10.1086/114227},
  urldate = {2026-07-03}
}

@article{Neugent+18,
  title = {A {{Modern Search}} for {{Wolf-Rayet Stars}} in the {{Magellanic Clouds}}. {{IV}}. {{A Final Census}}*},
  author = {Neugent, Kathryn F. and Massey, Philip and Morrell, Nidia},
  year = 2018,
  month = aug,
  journal = {ApJ},
  volume = {863},
  number = {2},
  pages = {181},
  publisher = {The American Astronomical Society},
  issn = {0004-637X},
  doi = {10.3847/1538-4357/aad17d},
  urldate = {2024-10-11}
}

@article{Nugis+00,
  title = {Mass-Loss Rates of Wolf-Rayet Stars as a Function of Stellar Parameters},
  author = {Nugis, T. and Lamers, H. J. G. L. M.},
  year = 2000,
  month = aug,
  journal = {A\&A},
  volume = {360},
  pages = {227--244},
  issn = {0004-6361},
  urldate = {2025-04-15}
}

@article{Nuijten+25,
  title = {{{WR}}+{{O}} Binaries as Probes of the First Phase of Mass Transfer},
  author = {Nuijten, Marit and Nelemans, Gijs},
  year = 2025,
  month = mar,
  journal = {A\&A},
  volume = {695},
  pages = {A117},
  publisher = {EDP Sciences},
  issn = {0004-6361, 1432-0746},
  doi = {10.1051/0004-6361/202451564},
  urldate = {2025-05-20},
  copyright = {\copyright{} The Authors 2025}
}

@article{Paczynski+67,
  title = {Evolution of {{Close Binaries}}. {{V}}. {{The Evolution}} of {{Massive Binaries}} and the {{Formation}} of the {{Wolf-Rayet Stars}}},
  author = {Paczy{\'n}ski, B.},
  year = 1967,
  month = jan,
  journal = {Acta Astronomica},
  volume = {17},
  pages = {355},
  issn = {0001-5237},
  urldate = {2024-10-31}
}

@article{Parkosidis+26,
  title = {Eccentricity as a Probe of Mass-Transfer Physics. {{Eccentric}} Mass Transfer as a Solution to the Wide Eccentric Binary Problem},
  author = {Parkosidis, A. and Toonen, S. and Laplace, E. and Schaffenroth, V.},
  year = 2026,
  month = jun,
  publisher = {arXiv},
  doi = {10.48550/arXiv.2606.09464},
  urldate = {2026-07-10}
}

@article{Pauli+22,
  title = {A Synthetic Population of Wolf-Rayet Stars in the {{LMC}} Based on Detailed Single and Binary Star Evolution Models},
  author = {Pauli, D. and Langer, N. and {Aguilera-Dena}, D. R. and Wang, C. and Marchant, P.},
  year = 2022,
  month = nov,
  journal = {A\&A},
  volume = {667},
  pages = {A58},
  issn = {0004-6361},
  doi = {10.1051/0004-6361/202243965},
  urldate = {2023-07-26}
}

@article{Pauli+26,
  title = {The Drastic Impact of {{Eddington-limit}} Induced Mass Ejections on Massive Star Populations},
  author = {Pauli, D. and Langer, N. and Schootemeijer, A. and Marchant, P. and Jin, H. and Ercolino, A. and Picco, A. and Willcox, R. and Sana, H.},
  year = 2026,
  month = jan,
  number = {arXiv:2601.08822},
  eprint = {2601.08822},
  primaryclass = {astro-ph},
  publisher = {arXiv},
  doi = {10.48550/arXiv.2601.08822},
  urldate = {2026-01-16},
  archiveprefix = {arXiv}
}

@article{Paxton+11,
  type = {Solar and {{Stellar Astrophysics}}; {{Instrumentation}} and {{Methods}} for {{Astrophysics}}},
  title = {Modules for Experiments in Stellar Astrophysics (Mesa)},
  author = {Paxton, Bill and Bildsten, Lars and Dotter, Aaron and Herwig, Falk and Lesaffre, Pierre and Timmes, Frank},
  year = 2011,
  month = jan,
  journal = {ApJS},
  volume = {192},
  number = {1},
  eprint = {1009.1622},
  pages = {3},
  issn = {0067-0049},
  doi = {10.1088/0067-0049/192/1/3},
  urldate = {2016-02-27},
  archiveprefix = {arXiv}
}

@article{Paxton+13,
  type = {Solar and {{Stellar Astrophysics}}; {{Instrumentation}} and {{Methods}} for {{Astrophysics}}},
  title = {Modules for Experiments in Stellar Astrophysics (Mesa): Planets, Oscillations, Rotation, and Massive Stars},
  author = {Paxton, Bill and Cantiello, Matteo and Arras, Phil and Bildsten, Lars and Brown, Edward F. and Dotter, Aaron and Mankovich, Christopher and Montgomery, M. H. and Stello, Dennis and Timmes, F. X. and Townsend, Richard},
  year = 2013,
  month = sep,
  journal = {ApJS},
  volume = {208},
  number = {1},
  eprint = {1301.0319},
  pages = {4},
  issn = {0067-0049},
  doi = {10.1088/0067-0049/208/1/4},
  urldate = {2016-05-10},
  archiveprefix = {arXiv}
}

@article{Paxton+15,
  title = {Modules for Experiments in Stellar Astrophysics ({{MESA}}): Binaries, Pulsations, and Explosions},
  author = {Paxton, Bill and Marchant, Pablo and Schwab, Josiah and Bauer, Evan B. and Bildsten, Lars and Cantiello, Matteo and Dessart, Luc and Farmer, R. and Hu, H. and Langer, N. and Townsend, R. H. D. and Townsley, Dean M. and Timmes, F. X.},
  year = 2015,
  month = jun,
  journal = {ApJS},
  volume = {220},
  number = {1},
  eprint = {1506.03146},
  pages = {15},
  issn = {1538-4365},
  doi = {10.1088/0067-0049/220/1/15},
  urldate = {2016-05-10},
  archiveprefix = {arXiv}
}

@article{Paxton+18,
  title = {Modules for Experiments in Stellar Astrophysics ({{MESA}}): Convective Boundaries, Element Diffusion, and Massive Star Explosions},
  shorttitle = {Modules for Experiments in Stellar Astrophysics ({{MESA}})},
  author = {Paxton, Bill and Schwab, Josiah and Bauer, Evan B. and Bildsten, Lars and Blinnikov, Sergei and Duffell, Paul and Farmer, R. and Goldberg, Jared A. and Marchant, Pablo and Sorokina, Elena and Thoul, Anne and Townsend, Richard H. D. and Timmes, F. X.},
  year = 2018,
  month = feb,
  journal = {ApJS},
  volume = {234},
  pages = {34},
  issn = {0067-0049},
  doi = {10.3847/1538-4365/aaa5a8},
  urldate = {2022-08-17}
}

@article{Paxton+19,
  title = {Modules for Experiments in Stellar Astrophysics ({{MESA}}): Pulsating Variable Stars, Rotation, Convective Boundaries, and Energy Conservation},
  shorttitle = {Modules for Experiments in Stellar Astrophysics ({{MESA}})},
  author = {Paxton, Bill and Smolec, R. and Schwab, Josiah and Gautschy, A. and Bildsten, Lars and Cantiello, Matteo and Dotter, Aaron and Farmer, R. and Goldberg, Jared A. and Jermyn, Adam S. and Kanbur, S. M. and Marchant, Pablo and Thoul, Anne and Townsend, Richard H. D. and Wolf, William M. and Zhang, Michael and Timmes, F. X.},
  year = 2019,
  month = jul,
  journal = {ApJS},
  volume = {243},
  pages = {10},
  issn = {0067-0049},
  doi = {10.3847/1538-4365/ab2241},
  urldate = {2022-08-17}
}

@article{Petrovic+05a,
  title = {Constraining the Mass Transfer in Massive Binaries through Progenitor Evolution Models of {{Wolf-Rayet}}+{{O}} Binaries},
  author = {Petrovic, J. and Langer, N. and {van der Hucht}, K. A.},
  year = 2005,
  month = jun,
  journal = {A\&A},
  volume = {435},
  pages = {1013--1030},
  issn = {0004-6361},
  doi = {10.1051/0004-6361:20042368},
  urldate = {2024-10-16}
}

@article{Prestwich+07,
  title = {The Orbital Period of the Wolf-Rayet Binary {{IC}} 10 {{X-1}}: Dynamic Evidence That the Compact Object Is a Black Hole},
  shorttitle = {The Orbital Period of the Wolf-Rayet Binary {{IC}} 10 {{X-1}}},
  author = {Prestwich, A. H. and Kilgard, R. and Crowther, P. A. and Carpano, S. and Pollock, A. M. T. and Zezas, A. and Saar, S. H. and Roberts, T. P. and Ward, M. J.},
  year = 2007,
  month = nov,
  journal = {ApJ},
  volume = {669},
  pages = {L21-L24},
  publisher = {IOP},
  issn = {0004-637X},
  doi = {10.1086/523755},
  urldate = {2026-03-09}
}

@incollection{Reimers+75,
  title = {Circumstellar {{Envelopes}} and {{Mass Loss}} of {{Red Giant Stars}}},
  booktitle = {Problems in Stellar Atmospheres and Envelopes},
  author = {Reimers, Dieter},
  editor = {Baschek, Bodo and Kegel, Wilhelm H. and Traving, Gerhard},
  year = 1975,
  pages = {229--256},
  publisher = {Springer},
  address = {Berlin, Heidelberg},
  doi = {10.1007/978-3-642-80919-4_8},
  urldate = {2026-07-03},
  isbn = {978-3-642-80919-4}
}

@article{Richardson+24,
  title = {Visual Orbits of Wolf-rayet Stars. {{I}}. {{The}} Orbit of the Dust-Producing Wolf-rayet Binary {{WR}} 137 Measured with the {{CHARA}} Array},
  author = {Richardson, Noel D. and Schaefer, Gail H. and Eldridge, Jan J. and Spejcher, Rebecca and Holdsworth, Amanda and Lau, Ryan M. and Monnier, John D. and Moffat, Anthony F. J. and Weigelt, Gerd and Williams, Peredur M. and Kraus, Stefan and Le Bouquin, Jean-Baptiste and Anugu, Narsireddy and Chhabra, Sorabh and Codron, Isabelle and Ennis, Jacob and Gardner, Tyler and Gutierrez, Mayra and Ibrahim, Noura and Labdon, Aaron and Lanthermann, Cyprien and Setterholm, Benjamin R.},
  year = 2024,
  month = dec,
  journal = {ApJ},
  volume = {977},
  number = {1},
  pages = {78},
  issn = {0004-637X},
  doi = {10.3847/1538-4357/ad8d5c},
  urldate = {2026-07-13}
}

@article{Rocha+24,
  title = {To {{Be}} or {{Not To Be}}: {{The Role}} of {{Rotation}} in {{Modeling Galactic Be X-Ray Binaries}}},
  shorttitle = {To {{Be}} or {{Not To Be}}},
  author = {Rocha, Kyle Akira and Kalogera, Vicky and Doctor, Zoheyr and Andrews, Jeff J. and Sun, Meng and Gossage, Seth and Bavera, Simone S. and Fragos, Tassos and Kovlakas, Konstantinos and Kruckow, Matthias U. and Misra, Devina and Srivastava, Philipp M. and Xing, Zepei and Zapartas, Emmanouil},
  year = 2024,
  month = aug,
  journal = {ApJ},
  volume = {971},
  pages = {133},
  publisher = {IOP},
  issn = {0004-637X},
  doi = {10.3847/1538-4357/ad5955},
  urldate = {2024-10-10}
}

@article{Sana+12,
  title = {Binary Interaction Dominates the Evolution of Massive Stars},
  author = {Sana, H and {de Mink}, S E and {de Koter}, A and Langer, N and Evans, C J and Gieles, M and Gosset, E and Izzard, R G and Le Bouquin, J-B and Schneider, F R N},
  year = 2012,
  month = jul,
  journal = {Sci. (N. Y. N.Y.)},
  volume = {337},
  number = {6093},
  eprint = {1207.6397},
  pages = {444--6},
  issn = {1095-9203},
  doi = {10.1126/science.1223344},
  urldate = {2016-03-23},
  archiveprefix = {arXiv},
  pmid = {22837522}
}

@article{Sana+13,
  title = {The {{VLT-FLAMES Tarantula Survey}}. {{VIII}}. {{Multiplicity}} Properties of the {{O-type}} Star Population},
  author = {Sana, H. and {de Koter}, A. and {de Mink}, S. E. and Dunstall, P. R. and Evans, C. J. and {H{\'e}nault-Brunet}, V. and Ma{\'i}z Apell{\'a}niz, J. and {Ram{\'i}rez-Agudelo}, O. H. and Taylor, W. D. and Walborn, N. R. and Clark, J. S. and Crowther, P. A. and Herrero, A. and Gieles, M. and Langer, N. and Lennon, D. J. and Vink, J. S.},
  year = 2013,
  month = feb,
  journal = {A\&A},
  volume = {550},
  pages = {A107},
  issn = {0004-6361},
  doi = {10.1051/0004-6361/201219621},
  urldate = {2022-03-30}
}

@article{Sana+25,
  title = {A High Fraction of Close Massive Binary Stars at Low Metallicity},
  author = {Sana, H. and Shenar, T. and Bodensteiner, J. and Britavskiy, N. and Langer, N. and Lennon, D. J. and Mahy, L. and Mandel, I. and {de Mink}, S. E. and Patrick, L. R. and Villase{\~n}or, J. I. and Dirickx, M. and {Abdul-Masih}, M. and Almeida, L. A. and Backs, F. and Berlanas, S. R. and {Bernini-Peron}, M. and Bowman, D. M. and Bronner, V. A. and Crowther, P. A. and Deshmukh, K. and Evans, C. J. and Fabry, M. and Gieles, M. and Gilkis, A. and {Gonz{\'a}lez-Tor{\`a}}, G. and Gr{\"a}fener, G. and G{\"o}tberg, Y. and Hawcroft, C. and {H{\'e}nault-Brunet}, V. and Herrero, A. and Holgado, G. and Izzard, R. G. and {de Koter}, A. and Janssens, S. and Johnston, C. and Josiek, J. and Justham, S. and Kalari, V. M. and Klencki, J. and Kub{\'a}t, J. and Kub{\'a}tov{\'a}, B. and Lefever, R. R. and {van Loon}, J. {\relax Th}. and Ludwig, B. and Mackey, J. and Ma{\'i}z Apell{\'a}niz, J. and Maravelias, G. and Marchant, P. and Mazeh, T. and Menon, A. and Moe, M. and Najarro, F. and Oskinova, L. M. and Ovadia, R. and Pauli, D. and Pawlak, M. and Ramachandran, V. and Renzo, M. and Rocha, D. F. and Sander, A. A. C. and Schneider, F. R. N. and Schootemeijer, A. and Sch{\"o}sser, E. C. and Sch{\"u}rmann, C. and Sen, K. and Shahaf, S. and {Sim{\'o}n-D{\'i}az}, S. and {van Son}, L. A. C. and Stoop, M. and Toonen, S. and Tramper, F. and Valli, R. and {Vigna-G{\'o}mez}, A. and Vink, J. S. and Wang, C. and Willcox, R.},
  year = 2025,
  month = sep,
  journal = {Nat. Astron},
  volume = {9},
  pages = {1337--1346},
  issn = {2397-3366},
  doi = {10.1038/s41550-025-02610-x},
  urldate = {2026-04-30}
}

@article{Schmutz+89,
  title = {Spectral Analysis of 30 Wolf-Rayet Stars},
  author = {Schmutz, W. and Hamann, W.-R. and Wessolowski, U.},
  year = 1989,
  month = feb,
  journal = {A\&A},
  volume = {210},
  pages = {236--248},
  publisher = {EDP},
  issn = {0004-6361},
  urldate = {2026-07-03}
}

@article{Schnurr+08,
  title = {A Spectroscopic Survey of {{WNL}} Stars in the {{Large Magellanic Cloud}}: General Properties and Binary Status},
  shorttitle = {A Spectroscopic Survey of {{WNL}} Stars in the {{Large Magellanic Cloud}}},
  author = {Schnurr, O. and Moffat, A. F. J. and {St-Louis}, N. and Morrell, N. I. and Guerrero, M. A.},
  year = 2008,
  month = sep,
  journal = {MNRAS},
  volume = {389},
  pages = {806--828},
  publisher = {OUP},
  issn = {0035-8711},
  doi = {10.1111/j.1365-2966.2008.13584.x},
  urldate = {2026-07-03}
}

@article{Schootemeijer+18,
  title = {Wolf-{{Rayet}} Stars in the {{Small Magellanic Cloud}} as Testbed for Massive Star Evolution},
  author = {Schootemeijer, A. and Langer, N.},
  year = 2018,
  month = mar,
  journal = {A\&A},
  volume = {611},
  pages = {A75},
  issn = {0004-6361},
  doi = {10.1051/0004-6361/201731895},
  urldate = {2024-10-20}
}

@article{Schootemeijer+24,
  title = {An Absence of Binary Companions to {{Wolf-Rayet}} Stars in the {{Small Magellanic Cloud}}: {{Implications}} for Mass Loss and Black Hole Masses at Low Metallicities},
  shorttitle = {An Absence of Binary Companions to {{Wolf-Rayet}} Stars in the {{Small Magellanic Cloud}}},
  author = {Schootemeijer, A. and Shenar, T. and Langer, N. and Grin, N. and Sana, H. and Gr{\"a}fener, G. and Sch{\"u}rmann, C. and Wang, C. and Xu, X. -T.},
  year = 2024,
  month = sep,
  journal = {A\&A},
  volume = {689},
  pages = {A157},
  issn = {0004-6361},
  doi = {10.1051/0004-6361/202449978},
  urldate = {2024-09-22}
}

@article{Sen+22,
  ids = {sen_2022a},
  title = {Detailed Models of Interacting Short-Period Massive Binary Stars},
  author = {Sen, K. and Langer, N. and Marchant, P. and Menon, A. and {de Mink}, S. E. and Schootemeijer, A. and Sch{\"u}rmann, C. and Mahy, L. and Hastings, B. and Nathaniel, K. and Sana, H. and Wang, C. and Xu, X. T.},
  year = 2022,
  month = mar,
  journal = {A\&A},
  volume = {659},
  pages = {A98},
  publisher = {EDP Sciences},
  issn = {0004-6361},
  doi = {10.1051/0004-6361/202142574},
  urldate = {2023-05-02}
}

@article{Shara+16,
  title = {The Spin Rates of {{O}} Stars in {{WR}} + {{O}} Binaries. {{I}}. {{Motivation}}, Methodology and First Results from {{SALT}}},
  author = {Shara, Michael M. and Crawford, Steven M. and Vanbeveren, Dany and Moffat, Anthony F. J. and Zurek, David and Crause, Lisa},
  year = 2016,
  month = aug,
  number = {arXiv:1511.00046},
  eprint = {1511.00046},
  primaryclass = {astro-ph.SR},
  publisher = {arXiv},
  doi = {10.48550/arXiv.1511.00046},
  urldate = {2026-07-10},
  archiveprefix = {arXiv}
}

@article{Shenar+16,
  title = {Wolf-Rayet Stars in the Small Magellanic Cloud. {{II}}. {{Analysis}} of the Binaries},
  author = {Shenar, T. and Hainich, R. and Todt, H. and Sander, A. and Hamann, W.-R. and Moffat, A. F. J. and Eldridge, J. J. and Pablo, H. and Oskinova, L. M. and Richardson, N. D.},
  year = 2016,
  month = jun,
  journal = {A\&A},
  volume = {591},
  pages = {A22},
  publisher = {EDP},
  issn = {0004-6361},
  doi = {10.1051/0004-6361/201527916},
  urldate = {2026-02-23}
}

@article{Shenar+19,
  ids = {shenar_2019a,shenar_2019b},
  title = {The {{Wolf}}--{{Rayet}} Binaries of the Nitrogen Sequence in the {{Large Magellanic Cloud}} - {{Spectroscopy}}, Orbital Analysis, Formation, and Evolution},
  author = {Shenar, T. and Sablowski, D. P. and Hainich, R. and Todt, H. and Moffat, A. F. J. and Oskinova, L. M. and Ramachandran, V. and Sana, H. and Sander, A. a. C. and Schnurr, O. and {St-Louis}, N. and Vanbeveren, D. and G{\"o}tberg, Y. and Hamann, W.-R.},
  year = 2019,
  month = jul,
  journal = {A\&A},
  volume = {627},
  pages = {A151},
  publisher = {EDP Sciences},
  issn = {0004-6361, 1432-0746},
  doi = {10.1051/0004-6361/201935684},
  urldate = {2023-05-11},
  copyright = {\copyright{} ESO 2019}
}

@article{Shenar+20,
  title = {Why Binary Interaction Does Not Necessarily Dominate the Formation of {{Wolf-Rayet}} Stars at Low Metallicity},
  author = {Shenar, T. and Gilkis, A. and Vink, J. S. and Sana, H. and Sander, A. A. C.},
  year = 2020,
  month = feb,
  journal = {A\&A},
  volume = {634},
  pages = {A79},
  issn = {0004-6361},
  doi = {10.1051/0004-6361/201936948},
  urldate = {2023-05-18}
}

@book{Shenar+26,
  title = {Wolf-{{Rayet}} Stars},
  author = {Shenar, Tomer},
  year = 2026,
  month = jan,
  series = {Encyclopedia of {{Astrophysics}}},
  volume = {2},
  publisher = {Elsevier},
  address = {eprint: arXiv:2410.04436},
  doi = {10.1016/B978-0-443-21439-4.00031-6},
  urldate = {2026-08-11}
}

@article{Singh+02,
  title = {New Measurements of Orbital Period Change in Cygnus {{X-3}}},
  author = {Singh, N. S. and Naik, S. and Paul, B. and Agrawal, P. C. and Rao, A. R. and Singh, K. Y.},
  year = 2002,
  month = sep,
  journal = {A\&A},
  volume = {392},
  number = {1},
  pages = {161--167},
  issn = {0004-6361, 1432-0746},
  doi = {10.1051/0004-6361:20020923},
  urldate = {2026-07-10}
}

@article{Smith+14,
  title = {Mass Loss: Its Effect on the Evolution and Fate of High-Mass Stars},
  shorttitle = {Mass Loss},
  author = {Smith, Nathan},
  year = 2014,
  month = aug,
  journal = {ARA\&A},
  volume = {52},
  pages = {487--528},
  issn = {0066-4146},
  doi = {10.1146/annurev-astro-081913-040025},
  urldate = {2026-07-03}
}

@article{Smith+96,
  title = {A Three-Dimensional Classification for {{WN}} Stars},
  author = {Smith, Lindsey F. and Shara, Michael M. and Moffat, Anthony F. J.},
  year = 1996,
  month = jul,
  journal = {MNRAS},
  volume = {281},
  pages = {163--191},
  publisher = {OUP},
  issn = {0035-8711},
  doi = {10.1093/mnras/281.1.163},
  urldate = {2026-07-03}
}

@article{Stanway+20a,
  title = {Binary Fraction Indicators in Resolved Stellar Populations and Supernova-Type Ratios},
  author = {Stanway, E. R. and Eldridge, J. J. and Chrimes, A. A.},
  year = 2020,
  month = sep,
  journal = {MNRAS},
  volume = {497},
  pages = {2201--2212},
  publisher = {OUP},
  issn = {0035-8711},
  doi = {10.1093/mnras/staa2089},
  urldate = {2025-11-10}
}

@article{Vanbeveren+80,
  title = {On the Binary Frequency Distribution and Evolution of {{Wolf-Rayet}} Stars .},
  author = {Vanbeveren, D. and Conti, P. S.},
  year = 1980,
  month = aug,
  journal = {A\&A},
  volume = {88},
  pages = {230--239},
  publisher = {EDP},
  issn = {0004-6361},
  urldate = {2026-07-03}
}

@article{Vanbeveren+98a,
  title = {The {{WR}} and {{O-type}} Star Population Predicted by Massive Star Evolutionary Synthesis},
  author = {Vanbeveren, D. and De Donder, E. and Van Bever, J. and Van Rensbergen, W. and De Loore, C.},
  year = 1998,
  month = nov,
  journal = {New Astron.},
  volume = {3},
  pages = {443--492},
  publisher = {Elsevier},
  issn = {1384-1076},
  doi = {10.1016/S1384-1076(98)00020-7},
  urldate = {2026-07-10}
}

@article{vandenHeuvel+17,
  title = {Forming Short-Period Wolf--Rayet {{X-ray}} Binaries and Double Black Holes through Stable Mass Transfer},
  author = {{van den Heuvel}, E. P. J. and Portegies Zwart, S. F. and {de Mink}, S. E.},
  year = 2017,
  month = nov,
  journal = {MNRAS},
  volume = {471},
  number = {4},
  pages = {4256--4264},
  issn = {0035-8711},
  doi = {10.1093/mnras/stx1430},
  urldate = {2021-07-06}
}

@article{vanderHucht+01,
  title = {The {{VIIth}} Catalogue of Galactic Wolf--Rayet Stars},
  author = {{van der Hucht}, Karel A.},
  year = 2001,
  month = feb,
  journal = {New Astron. Rev.},
  volume = {45},
  number = {3},
  pages = {135--232},
  issn = {1387-6473},
  doi = {10.1016/S1387-6473(00)00112-3},
  urldate = {2026-01-21}
}

@article{vanSon+26,
  title = {Ongoing and Post-Mass-Transfer Binaries: A Living Catalog and Unified Review of Binary Mass Transfer Products},
  shorttitle = {Ongoing and Post-Mass-Transfer Binaries},
  author = {{van Son}, Lieke A. C. and Yamaguchi, Natsuko and Nagarajan, Pranav and Shenar, Tomer and Sen, Koushik and Laroche, Alexander and Leiner, Emily M. and Sana, Hugues and Pols, Onno R.},
  year = 2026,
  month = may,
  number = {arXiv:2605.31290},
  eprint = {2605.31290},
  primaryclass = {astro-ph.SR},
  publisher = {arXiv},
  doi = {10.48550/arXiv.2605.31290},
  urldate = {2026-06-01},
  archiveprefix = {arXiv}
}

@article{Veen+98,
  title = {A Second Dust Episode of the {{Wolf-Rayet}} System {{WR}} 19: Another Long-Period {{WC}}+{{O}} Colliding-Wind Binary},
  shorttitle = {A Second Dust Episode of the {{Wolf-Rayet}} System {{WR}} 19},
  author = {Veen, P. M. and {van der Hucht}, K. A. and Williams, P. M. and Catchpole, R. M. and Duijsens, M. F. J. and Glass, I. S. and Setia Gunawan, D. Y. A.},
  year = 1998,
  month = nov,
  journal = {A\&A},
  volume = {339},
  pages = {L45-L48},
  publisher = {EDP},
  issn = {0004-6361},
  urldate = {2026-07-03}
}

@article{Vink+00,
  title = {New Theoretical Mass-Loss Rates of {{O}} and {{B}} Stars},
  author = {Vink, Jorick S. and {de Koter}, Alex and Lamers, Henny J. G. L. M.},
  year = 2000,
  month = aug,
  journal = {A\&A},
  volume = {362},
  eprint = {astro-ph/0008183},
  pages = {295--309},
  urldate = {2016-05-10},
  archiveprefix = {arXiv}
}

@article{Vink+01,
  title = {Mass-Loss Predictions for {{O}} and {{B}} Stars as a Function of Metallicity},
  author = {Vink, Jorick S. and {de Koter}, Alex and Lamers, Henny J. G. L. M.},
  year = 2001,
  month = jan,
  journal = {A\&A},
  volume = {369},
  eprint = {astro-ph/0101509},
  pages = {574--588},
  issn = {0004-6361},
  doi = {10.1051/0004-6361:20010127},
  urldate = {2016-05-10},
  archiveprefix = {arXiv}
}

@article{Vink+05,
  title = {On the Metallicity Dependence of Wolf-Rayet Winds},
  author = {Vink, Jorick S. and De Koter, A.},
  year = 2005,
  month = nov,
  journal = {A\&A},
  volume = {442},
  number = {2},
  pages = {587--596},
  issn = {0004-6361, 1432-0746},
  doi = {10.1051/0004-6361:20052862},
  urldate = {2024-11-11}
}

@article{Webbink+84,
  ids = {webbink_1984a},
  title = {Double White Dwarfs as Progenitors of {{R}} Coronae Borealis Stars and Type {{I}} Supernovae},
  author = {Webbink, R. F.},
  year = 1984,
  month = feb,
  journal = {ApJ},
  volume = {277},
  pages = {355--360},
  doi = {10.1086/161701},
  urldate = {2020-08-13}
}

@article{Wellstein+99,
  title = {Implications of Massive Close Binaries for Black Hole Formation and Supernovae},
  author = {Wellstein, S. and Langer, N.},
  year = 1999,
  month = aug,
  journal = {ArXivastro-Ph9904256},
  eprint = {astro-ph/9904256},
  urldate = {2021-02-22},
  archiveprefix = {arXiv}
}

@article{Williams+19,
  title = {Variable Dust Emission by {{WC}} Type Wolf-Rayet Stars Observed in the {{NEOWISE-R}} Survey},
  author = {Williams, P. M.},
  year = 2019,
  month = sep,
  journal = {MNRAS},
  volume = {488},
  pages = {1282--1300},
  publisher = {OUP},
  issn = {0035-8711},
  doi = {10.1093/mnras/stz1784},
  urldate = {2026-07-03}
}

@article{Wolf+67,
  title = {Spectroscopie Stellaire},
  author = {Wolf, C. J. E. and Rayet, G.},
  year = 1867,
  month = jan,
  journal = {Acad. Sci. Paris C. R.},
  volume = {65},
  pages = {292--296},
  urldate = {2026-07-03}
}

@article{Wong+14,
  title = {Understanding {{Compact Object Formation}} and {{Natal Kicks}}. {{IV}}. {{The Case}} of {{IC}} 10 {{X-1}}},
  author = {Wong, Tsing-Wai and Valsecchi, Francesca and Ansari, Asna and Fragos, Tassos and Glebbeek, Evert and Kalogera, Vassiliki and McClintock, Jeffrey},
  year = 2014,
  month = aug,
  journal = {ApJ},
  volume = {790},
  pages = {119},
  publisher = {IOP},
  issn = {0004-637X},
  doi = {10.1088/0004-637X/790/2/119},
  urldate = {2026-06-23}
}

@article{Xu+25,
  title = {Populations of Evolved Massive Binary Stars in the Small Magellanic Cloud {{I}}: Predictions from Detailed Evolution Models},
  shorttitle = {Populations of Evolved Massive Binary Stars in the Small Magellanic Cloud {{I}}},
  author = {Xu, X. -T. and Sch{\"u}rmann, C. and Langer, N. and Wang, C. and Schootemeijer, A. and Shenar, T. and Ercolino, A. and Haberl, F. and Hastings, B. and Jin, H. and Kramer, M. and Lennon, D. and Marchant, P. and Sen, K. and Tauris, T. M. and {de Mink}, S. E.},
  year = 2025,
  month = mar,
  publisher = {arXiv},
  doi = {10.48550/arXiv.2503.23876},
  urldate = {2025-04-15}
}

@article{Zapartas+25,
  title = {The Demographics of Binary Companions to Stripped-Envelope Supernovae: Confronting Observations with Population Synthesis},
  shorttitle = {The Demographics of Binary Companions to Stripped-Envelope Supernovae},
  author = {Zapartas, E. and Fox, O. D. and Su, J. and Souropanis, D. and Drout, M. R. and Rocha, K. A. and {van Dyk}, S. D. and Williams, B. F. and Briel, M. and Renzo, M. and Andrews, J. J. and Fragos, T. and Gossage, S. and Kruckow, M. U. and Liotine, C. and Ryder, S. D. and Srivastava, P. M. and Teng, E.},
  year = 2025,
  month = aug,
  publisher = {arXiv},
  doi = {10.48550/arXiv.2508.12677},
  urldate = {2025-11-12}
}

@article{Zapartas+25a,
  title = {The Effect of Mass Loss in Models of Red Supergiants in the Small Magellanic Cloud},
  author = {Zapartas, E. and {de Wit}, S. and Antoniadis, K. and {Mu{\~n}oz-Sanchez}, G. and Souropanis, D. and Bonanos, A. Z. and Maravelias, G. and Kovlakas, K. and Kruckow, M. U. and Fragos, T. and Andrews, J. J. and Bavera, S. S. and Briel, M. and Gossage, S. and Kasdagli, E. and Rocha, K. A. and Sun, M. and Srivastava, P. M. and Xing, Z.},
  year = 2025,
  month = may,
  journal = {A\&A},
  volume = {697},
  pages = {A167},
  publisher = {EDP},
  issn = {0004-6361},
  doi = {10.1051/0004-6361/202452401},
  urldate = {2026-04-16}
}

@article{Asplund+09,
  title = {The Chemical Composition of the {{Sun}}},
  author = {Asplund, Martin and Grevesse, Nicolas and Sauval, A. Jacques and Scott, Pat},
  year = 2009,
  month = sep,
  journal = {ARA\&A},
  volume = {47},
  number = {1},
  eprint = {0909.0948},
  pages = {481--522},
  issn = {0066-4146, 1545-4282},
  doi = {10.1146/annurev.astro.46.060407.145222},
  urldate = {2020-10-07},
  archiveprefix = {arXiv}
}

@article{Castor+75,
  title = {Radiation-Driven Winds in {{Of}} Stars.},
  author = {Castor, J. I. and Abbott, D. C. and Klein, R. I.},
  year = 1975,
  month = jan,
  journal = {ApJ},
  volume = {195},
  pages = {157--174},
  publisher = {IOP},
  issn = {0004-637X},
  doi = {10.1086/153315},
  urldate = {2026-08-14}
}

@article{Grafener+11,
  title = {The {{Eddington}} Factor as the Key to Understand the Winds of the Most Massive Stars. {{Evidence}} for a {{$\Gamma$-dependence}} of {{Wolf-Rayet}} Type Mass Loss},
  author = {Gr{\"a}fener, G. and Vink, J. S. and {de Koter}, A. and Langer, N.},
  year = 2011,
  month = nov,
  journal = {A\&A},
  volume = {535},
  pages = {A56},
  issn = {0004-6361},
  doi = {10.1051/0004-6361/201116701},
  urldate = {2026-08-14}
}

@article{Sen+23,
  title = {Reverse {{Algols}} and Hydrogen-Rich {{Wolf-Rayet}} Stars from Very Massive Binaries},
  author = {Sen, K. and Langer, N. and Pauli, D. and Gr{\"a}fener, G. and Schootemeijer, A. and Sana, H. and Shenar, T. and Mahy, L. and Wang, C.},
  year = 2023,
  month = apr,
  journal = {A\&A},
  volume = {672},
  pages = {A198},
  issn = {0004-6361},
  doi = {10.1051/0004-6361/202245378},
  urldate = {2023-07-26}
}

@article{Bavera+21,
  ids = {bavera_2021c},
  title = {The Impact of Mass-Transfer Physics on the Observable Properties of Field Binary Black Hole Populations},
  author = {Bavera, Simone S. and Fragos, Tassos and Zevin, Michael and Berry, Christopher P. L. and Marchant, Pablo and Andrews, Jeff J. and Coughlin, Scott and Dotter, Aaron and Kovlakas, Konstantinos and Misra, Devina and {Serra-Perez}, Juan G. and Qin, Ying and Rocha, Kyle A. and {Rom{\'a}n-Garza}, Jaime and Tran, Nam H. and Zapartas, Emmanouil},
  year = 2021,
  month = mar,
  journal = {A\&A},
  volume = {647},
  pages = {A153},
  issn = {0004-6361},
  doi = {10.1051/0004-6361/202039804},
  urldate = {2022-01-10}
}

@article{Disberg+25,
  title = {The Kick Velocity Distribution of Isolated Neutron Stars},
  author = {Disberg, Paul and Mandel, Ilya},
  year = 2025,
  month = aug,
  journal = {ApJ},
  volume = {989},
  pages = {L8},
  publisher = {IOP},
  issn = {0004-637X},
  doi = {10.3847/2041-8213/adf286},
  urldate = {2026-04-15}
}

@article{Chattaraj+26a,
  title = {Double {{Neutron Star Delay Times Across Cosmic Metallicities}}: {{The Role}} of {{Helium Star Progenitors}}},
  shorttitle = {Double {{Neutron Star Delay Times Across Cosmic Metallicities}}},
  author = {Chattaraj, Abhishek and Andrews, Jeff J. and Briel, Max and Fragos, Tassos and Gossage, Seth and Kalogera, Vicky and Srivastava, Philipp M. and Teng, Elizabeth},
  year = 2026,
  month = jul,
  journal = {ApJ},
  volume = {1006},
  pages = {5},
  publisher = {IOP},
  issn = {0004-637X},
  doi = {10.3847/1538-4357/ae7a66},
  urldate = {2026-08-20}
}


\begin{appendix}

\section{Metallicity evolution with $\tau=0.5$} \label{app:optical_depth_boundary}

We adopt the optical depth boundary calibrated by \citet{Aguilera-Dena+22}. However, more recent works have picked lower $\tau$ thresholds to more closely reproduce the observed WR population \citep[see for example,][]{Pauli+26}. Given the uncertainty in the optical depth boundary, we additionally compute populations using $\tau=0.5$, which increases the number of systems in the WR population. As shown in Section~\ref{sec:other_WR_selection}, this lower threshold does not introduce qualitative differences in the resulting formation channels or period distribution at $\Zsun$.

\begin{figure}
    \centering
    \includegraphics[width=\linewidth]{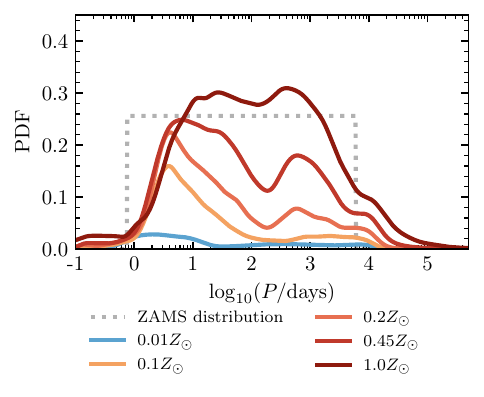}
    \caption{Same WR period distribution across metallicities as in Figure \ref{fig:metallicity_dependence} but with an optical depth selection criteria of $\tau=0.5$}
    \label{fig:metallicity_dependence_lower_tau}
\end{figure}

Figure~\ref{fig:metallicity_dependence_lower_tau} shows the WR binary period distribution across metallicity for $\tau=0.5$, which have the same overall trends as the $\tau=1.5$ populations presented in Section~\ref{sec:metallicity}. Long-period WR binaries remain absent at low metallicity, and short-period Case-A mass transfer continues to dominate WR binary formation. The main difference is that the short-period peak is broader at $\tau=0.5$ than at $\tau=1.5$.
his is likely due to less stripping being required to reach the wind strength and thus optical depth limit. As such, longer-period binaries through Case-A are able to experience a WR phase compared to a higher optical depth boundary.

\section{Table of relative contribution}

Table~\ref{tab:relative_contribution} contains the relative contributions of each formation channel at four metallicities as depicted in \ref{fig:formation_channels}.

\begin{table}[]
    \centering
    \begin{tabular}{c|c|c|c|c}
          & \multicolumn{4}{|c}{\textbf{Metallicity} ($\Zsun$)} \\
          \hline
       \textbf{Formation channel }  & 0.1 & 0.2 & 0.45 & 1 \\
         \hline
        SMT (S1)      & 47\% & 49\% & 45\% & 41\% \\
        SMT (S2)      & 16\% & 19\% & 19\% & 14\% \\
        No MT (S1)    & 20\% & 17\% & 16\% & 27\% \\
        No MT (S2)    & 7\%  & 7\%  & 10\% & 11\% \\
        Unstable (S1) & 5\%  & 3\%  & 2\% & 2\% \\
        Unstable (S2) & 4\%  & 4\%  & 5\% & 4\% \\
    \end{tabular}
    \caption{Relative contribution of each formation channel of WRs in binaries used in Figure \ref{fig:formation_channels}.}
    \label{tab:relative_contribution}
\end{table}

\section{HR diagram of 0.1\Zsun binaries}\label{app:HR_diagram_0.1Zsun}

\begin{figure}
    \centering
    \includegraphics[width=\linewidth]{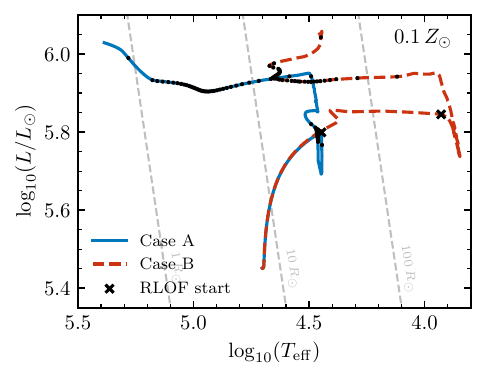}
    \caption{HR diagram of the $0.1\Zsun$ example models in Section \ref{sec:Case_A} (blue) and Section \ref{sec:Case_B} (red). The start of Roche lobe overflow (RLO) is marked with a black cross. After the mass transfer, we mark every 10000 years of the model with a black dot to indicate where the star spends most of its time.}
    \label{fig:HR_diagram_0.1Zsun}
\end{figure}

In Section \ref{sec:Case_B}, we showed how Case-B mass transfer does not lead to a WR phase at low metallicities.
With insufficient stellar wind mass-loss, the envelope is retained till the donor star reached core carbon depletion.
In Figure \ref{fig:HR_diagram_0.1Zsun}, we show the location of the $0.1\Zsun$ Case-A (blue) and Case-B (red) donor stars. Since the Case-B donor is not fully stripped, it does not reach a WR phase and instead stops in its movement to the hot (left) side of the HR diagram. The Case-A donor, on the other hand, reaches the hot side and spends most of its time after the mass transfer as a WR star.

\section{Initial period distribution} \label{app:initial_period_dist}

\begin{figure}
    \centering
    \includegraphics[width=\linewidth]{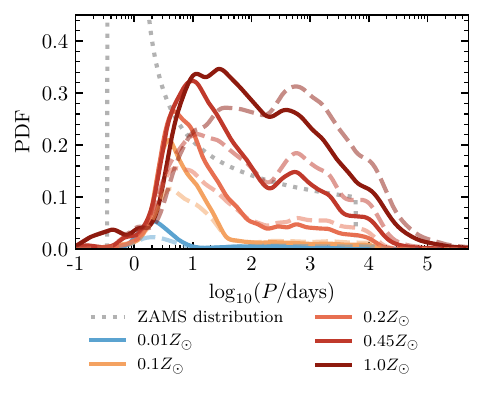}
    \caption{Same as Figure \ref{fig:metallicity_dependence} but the solid lines show the WR binary period distribution with a \citet{Sana+12} initial period distribution, which is shown as the dotted gray line. The original flat initial period distribution is shown as the dashed colored lines.}
    \label{fig:initial_period}
\end{figure}

The initial period distribution has a strong impact on the WR binary distribution by reweighting how much each initial binary contributed to the final period distribution. We have used an initially period distribution that is flat in $\log_{10}(P)$, but previous works have used a derivative of the \citet{Sana+12} initial period distribution to explore the WR binary population \citep{Pauli+22, Pauli+26, Xu+25}. Since this distribution prefers short-period systems, this decreases the contribution of Case-B mass transfer systems by reducing the number of initial systems undergoing Case-B mass transfer. 
Although recent observations suggest a flatter period distribution \citep[see references in][]{Sana+25}, we show the WR binary period distribution given a \citet{Sana+12} initial period distribution (following appendix A in \citet{Bavera+21}) in Figure \ref{fig:initial_period}. Comparing this distribution with solid lines against the flat in log-period distribution shows that the \citet{Sana+12} initial period distribution favors shorter period WR binaries, but that it does not remove the presence of long-period WR binaries at Milky Way (\Zsun) or LMC metallicity ($0.45\Zsun$). At SMC metallicity ($0.2\Zsun$), the shift in period distribution leads to a stronger preference for short-period WR binaries, but the decreasing efficiency of WR formation of Case-B mass transfer is still required to limit their contribution at low metallicity.

\end{appendix}
\end{document}